\documentclass[tighten,times]{aastex631}
\hypersetup{linkcolor=blue,citecolor=blue,filecolor=blue,urlcolor=magenta}
\newcommand{\um}{$\mu$m}

\definecolor{navyblue}{RGB}{0,50,250}

\definecolor{kostas}{rgb}{1.00, 0.00, 0.0}
\shorttitle{Extragalactic Magnetism with SOFIA - III: First Data Release}
\shortauthors{Lopez-Rodriguez et al.}

\begin{document}

\title{Extragalactic magnetism with SOFIA (SALSA Legacy Program) - III: \\
First data release and on-the-fly polarization mapping characterization\footnote{The SOFIA Legacy Program for Magnetic Fields in Galaxies (SALSA) provides a software repository at \url{https://github.com/galmagfields/hawc}, and publicly available data at \url{http://galmagfields.com/}}}

\correspondingauthor{Lopez-Rodriguez, E.}
\email{elopezrodriguez@stanford.edu}

\author[0000-0001-5357-6538]{Enrique Lopez-Rodriguez}
\affil{Kavli Institute for Particle Astrophysics \& Cosmology (KIPAC), Stanford University, Stanford, CA 94305, USA}
\author{Melanie Clarke}
\affil{SOFIA Science Center, NASA Ames Research Center, Moffett Field, CA 94035, USA}
\author{Sachin Shenoy}
\author{William Vacca}
\affil{SOFIA Science Center, NASA Ames Research Center, Moffett Field, CA 94035, USA}
\author{Simon Coude}
\affil{SOFIA Science Center, NASA Ames Research Center, Moffett Field, CA 94035, USA}
\author{Ryan Arneson}
\affil{SOFIA Science Center, NASA Ames Research Center, Moffett Field, CA 94035, USA}
\author{Peter Ashton}
\affil{SOFIA Science Center, NASA Ames Research Center, Moffett Field, CA 94035, USA}
\author{Sarah Eftekharzadeh}
\affil{SOFIA Science Center, NASA Ames Research Center, Moffett Field, CA 94035, USA}
\author{Rainer Beck}
\affil{Max-Planck-Institut f\"ur Radioastronomie, Auf dem H\"ugel 69, 53121 Bonn, Germany}

\author{John E. Beckman}
\affil{Instituto de Astrof\'isica de Canarias, C/ Via L\'actea s/n, E-38200 La Laguna, Tenerife, Spain}
\affil{Departamento de Astrof\'isica, Universidad de La Laguna, Avda. Astrof\'isico Fco. Sánchez s/n, E-38200 La Laguna, Tenerife, Spain}

\author[0000-0003-3249-4431]{Alejandro S. Borlaff}
\affil{NASA Ames Research Center, Moffett Field, CA 94035, USA}

\author[0000-0002-7633-3376]{Susan E. Clark}
\affil{Department of Physics, Stanford University, Stanford, CA 94305, USA}
\affil{Kavli Institute for Particle Astrophysics \& Cosmology (KIPAC), Stanford University, Stanford, CA 94305, USA}

\author[0000-0002-5782-9093]{Daniel A. Dale}
\affil{Department of Physics \& Astronomy, University of Wyoming, Laramie, WY 82071, USA}

\author[0000-0002-4059-9850]{Sergio Martin-Alvarez}
\affil{Institute of Astronomy and Kavli Institute for Cosmology, University of Cambridge, Madingley Road, Cambridge CB3 0HA, UK}

\author[0000-0002-4324-0034]{Evangelia Ntormousi}
\affil{Scuola Normale Superiore di Pisa, Piazza dei Cavalieri 7, 56126 Pisa, Italy}
\affil{Institute of Astrophysics, Foundation for Research and Technology-Hellas, Vasilika Vouton, GR-70013 Heraklion, Greece}

\author[0000-0001-8362-4094]{William T. Reach}
\affil{Universities Space Research Association, NASA Ames Research Center, Moffett Field, CA 94035, USA}

\author[0000-0001-6326-7069]{Julia Roman-Duval}
\affiliation{Space Telescope Science Institute 3700 San Martin Drive Baltimore, MD21218, USA}

\author[0000-0002-8831-2038]{Konstantinos Tassis}
\affil{Department of Physics \& ITCP, University of Crete, GR-70013, Heraklion, Greece}
\affil{Institute of Astrophysics, Foundation for Research and Technology-Hellas, Vasilika Vouton, GR-70013 Heraklion, Greece}

\author{Doyal A. Harper}
\affil{Department of Astronomy and Astrophysics University of Chicago, Chicago, IL 60637, USA}

\author{Pamela M. Marcum}
\affil{NASA Ames Research Center, Moffett Field, CA 94035, USA}


\begin{abstract} 
We describe the data processing of the Survey on extragALactic magnetiSm with SOFIA (SALSA Legacy Program). This first data release presents 33\% (51.34h out of 155.7h, including overheads) of the total awarded time taken from January 2020 to December 2021. Our observations were performed using the newly implemented on-the-fly mapping (OTFMAP) technique in the polarimetric mode. We present the pipeline steps to obtain homogeneously reduced high-level data products of polarimetric maps of galaxies for use in scientific analysis. Our approach has a general design and can be applied to sources smaller than the field-of-view of the HAWC+ array in any given band. We estimate that the OTFMAP polarimetric mode offers a reduction of observing overheads by a factor 2.34, and an improvement in sensitivity by a factor 1.80 when compared to previously obtained polarimetric observations using the chopping and nodding mode. The OTFMAP is a significant optimization of the polarimetric mode of HAWC+ as it ultimately reduces the cost of operations of SOFIA/HAWC+ by increasing the science collected per hour of observation up to an overall factor of 2.49. The OTFMAP polarimetric mode is the standard observing strategy of SALSA. The results and quantitative analysis of this first data release are presented in Papers IV and V of the series.
\end{abstract}


\section{Introduction} \label{sec:INT}

Infrared (IR) observations from ground-based and suborbital telescopes are dominated by atmospheric thermal emission and telescope background emission. Revealing the astrophysical sources requires observing strategies that remove or minimize the competing contributions of these background emissions. The general approach is to observe the combined emission from the source and the background with a detector, and then move the same detector to a part of the sky with only background emission. The offset between these two positions is performed on a timescale shorter than the background variation and the detector response. The subtraction of these two observations reveals the astrophysical source.

The chopping and nodding (C2N) strategy has been the most commonly used observing technique in the mid-IR (MIR; $8-30$ \um) wavelength range \citep[e.g.][and references therein]{Burtscher2020} and sometimes used in the far-IR (FIR; $30-300$ \um) wavelength range \citep[e.g.][]{Hildebrand2000}. The secondary mirror alternates between an on-axis and an adjacent off-axis position on a fast cadence. The angular separations (i.e. chop-throw) between on- and off-axis positions are usually within two to three fields-of-view (FOV) of the array in a given band, yielding $\sim5-30$\arcsec\ for MIR in 8-m class telescopes and up to $500$\arcsec\ for FIR wavelengths. The cadence of $<1$s facilitates subtraction of fast sky variations. This telescope movement is known as chopping. Each pair of on-axis and off-axis images are subtracted to eliminate the background emission from the science image. However, as the secondary mirror is not aligned with the optical axis of the telescope while observing the nearby sky location, a residual background emission (also named radiative offset) is still present in the subtracted image. To minimize the radiative offset, the telescope is moved on its optical axis every $\le60$s with the same, or different, amplitude and direction as the chopping configuration. A pair of subtracted chop-nod observations produces a positive image of the science object. 

The C2N observing strategy has several disadvantages. First, although the residual offset is subtracted, final reduced images still have some level of background that generates emission patterns and/or residual levels of radiation that are not entirely removed. These residuals may arise from radiation of structures in the dome, sky variations, detector response variations, and/or dependencies of radiation offset as the telescope rotates. Second, chop-throws are usually $\le 10-500$\arcsec, which can complicate the background subtraction for objects with extended diffuse thermal emission (e.g. the Galactic center, star-forming regions, molecular clouds). Finally, this observing mode spends approximately half of the total observing time pointing to a nearby sky location away from the science object, which makes the observations inefficient with large overheads.

The on-the-fly mapping (OTFMAP) strategy is an alternative observing mode in which the telescope on its optical axis scans the science object and surrounding areas. This technique is commonly used in the FIR and sub-mm wavelength ranges \citep[e.g.][]{Tegmark1997,Hildebrand2000,Reichertz2001,Weferling2002,Waskett2007,Chapin2013}. Single-dish radio telescopes also use this observing technique \citep{Haslam1974,Muller2017}. This technique has recently gained attention in the MIR wavelength range \citep[e.g.][]{Ohsawa2018} due to its prospects for the next generation of 30-m class telescopes. The telescope on its optical axis performs a scanning strategy \citep[e.g.][]{Kovacs2008b} described by a parametric curve and a mapping speed. The mapping speed is chosen such that the beam size (full width at half maximum, FWHM) of the instrument is oversampled at the sampling rate of the detector readout. For this configuration: a) the sampling rate has to be at least twice the maximum frequency of the astrophysical and/or sky signal (i.e. spatial and temporal Nyquist sampling), and b) the system noise must be comparable to or lower than the power spectrum of the sky at the desired spatial frequency to be recovered \citep[e.g.][]{Emerson1988}.  

The OTFMAP observing strategy presents advantages in terms of observing overheads as at any given time the scan is integrating over a deeper image. Also, the telescope can be moved to farther adjacent sky regions to reach `true' zero-level background emission. For bolometers, the signal is dominated by the $1/f$ noise, which is characterized by having higher amplitudes at longer timescales. This behavior results in signal drifts that can affect all spatial structures in the final image. Most of the efforts of the OTFMAP strategy focus on the map-making algorithms and signal filtering to recover diffuse and extended emission from the astrophysical object. A map-making algorithm is required to reconstruct the sky image while subtracting the sky and detector variation responses \citep[i.e.][]{Roussel2013,kovacs2006,Waskett2007,Patanchon2008,kovacs2008,Cantalupo2010,Chapin2013}. 

Chop-nod has been the standard observing strategy for the polarimetric mode of the High-resolution Airborne Wideband Camera-plus \citep[HAWC+;][]{Vaillancourt2007,Dowell2010,Harper2018} onboard the 2.7-m Stratospheric Observatory For Infrared Astronomy (SOFIA). The OTFMAP observing mode had previously only been used for total intensity observations \citep{Harper2018}. Recent observations of galaxies, Centaurus A \citep{ELR2021} and NGC~1097 \citep{ELR2021c}, and the molecular cloud Taurus/L1495 \citep{Li2021} have successfully used the OTFMAP polarimetric mode with HAWC+. However, a full characterization of this new observing mode for HAWC+ has not been presented until now.

This manuscript describes the data reduction scheme of the OTFMAP polarimetric mode of HAWC+ for sources smaller than the FOV in any given band. In Section \ref{sec:OBS}, we present the individual steps from raw data to high-level data products for use in scientific analysis. We present approaches for background subtraction, precipitable water vapour correction, and flux and polarization calibration. In Section \ref{sec:OTFvsC2N}, we show a quantitative comparison between C2N and OTFMAP polarization modes and estimate the sensitivities for the OTFMAP observing mode. First results of this data release will be presented in Paper IV and V of the series.


\section{Observations and Data Reduction} \label{sec:OBS}

\subsection{Observations}\label{subsec:OBS}

New multi-wavelength polarimetric observations of 10 galaxies were performed as part of the SOFIA Legacy Program\footnote{SOFIA Legacy Program on extragalactic magnetism can be found at \url{http://galmagfields.com/}} (ID: 08\_0012, PI: Lopez-Rodriguez, E. \& Mao, S. A.) using HAWC+ on the 2.7-m SOFIA telescope. This first data release presents 33\% (51.34h out of 155.7h, including overheads) of the total awarded time taken from January 2020 to December 2021. Table \ref{tab:OBS} shows the details of the new observations presented in this manuscript. In addition, we also include observations of Centaurus~A \citep{ELR2021} at 89 \um~and Circinus at 53, 89, and 214 \um~from program ID 07\_0032 (PI: Lopez-Rodriguez, E.), M51 \citep{Borlaff2021} at 154 \um~under programs 70\_0509 (Guaranteed Time Observations by the HAWC+ Team), 76\_0003 (Discretionary Director Time), and 08\_0260 (PI: Dowell, D.),  and NGC~1097 \citep{ELR2021c} at 89 and 154 \um~from program ID 07\_0034 (PI: Lopez-Rodriguez, E.). These additional observations comprised a total exposure time of 11.84h (including overheads) taken from 2017 to 2021. A total of 14 galaxies are presented in this data release.

All galaxies presented in this data release were homogeneously reduced using the steps presented below. For those galaxies already published, i.e. Centaurus~A and NGC~1097, we reprocessed these datasets to produce data products within a homogeneous reduction scheme. 

\startlongtable
\centerwidetable
\begin{deluxetable*}{lcccccccccccc}
\tablecaption{Summary of new OTFMAP polarimetric observations. \emph{Columns, from left to right:} a) Object name. b) Central wavelength of the band used for observations. c) Observation date (YYYYMMDD). d) Flight ID. e) Sea-level altitude during the observations (kft.). f) Speed of the scan (\arcsec/sec.). g) Amplitudes in elevation (EL) and cross-elevation (XEL) of the scan (\arcsec). h) Time per scan (s). i) Number of observation sets used (and rejected). j) On-source observation time (s).
\label{tab:OBS} 
}
\tablecolumns{7}
\tablewidth{0pt}
\tablehead{\colhead{Object}	&	\colhead{Band} &	\colhead{Date}	&	\colhead{Flight}	&	\colhead{Altitude}	&
\colhead{Scan Rate} & \colhead{Scan Amplitude } & \colhead{Scan Duration} & \colhead{\#Sets (bad)} & \colhead{Obs. Time} \\ 
 	&	& \colhead{(YYYYMMDD)}	&		&	\colhead{(kft)}	& \colhead{(\arcsec/sec.)} &  \colhead{ (EL $\times$ XEL; \arcsec)} & \colhead{(s)}  & & \colhead{(s)} \\
\colhead{(a)} & \colhead{(b)} & \colhead{(c)} & \colhead{(d)} & \colhead{(e)} & \colhead{(f)} & \colhead{(g)} & \colhead{(h)} & \colhead{(i)} &\colhead{(j)}
}
\startdata
Circinus$^{\star}$	& A (53 \um)	&	20190716	&	F596		&	42	&	100	&	50$\times$50	&	100	&	1(2)	&	400 \\
		&			&		&				&		&				&		&	&	Total time	& 400 (0.11h)	\\
		& C (89 \um)	&	20190716	&	F596		&	41-42	&	100	&	70$\times$70	&	100	&	2	& 800\\ 
		&			&			&			&				&		&	90$\times$90	&	100	&	6	& 2400 \\	
		&			&		&				&		&				&		&	&	Total time	& 3200 (0.89h)	\\
		& E (214 \um)	&	20190716	&	F596		&	43	&	100	&	120$\times$120	&	100, 60	& 2, 1	& 1040 \\
		&			&		&				&		&				&		&	&	Total time	& 1040 (0.29h)	\\				
\hline
M~82	& A (53 \um)	&	20210505	&	F726		&	41-42	&	100	&	80$\times$80	&	78, 60	&	1(2), 8(1)	& 2232 \\
		& 			&	20210513	&	F731		&	43	&	100	&	80$\times$80	&	78	&	11	&	3432		\\
		&			&		&				&		&				&		&	&	Total time	& 5664 (1.6h)	\\
		& C (89 \um)	&	20210518	&	F733		&	43	&	100	&	90$\times$90	&	78	&	11	&	3432	\\
		&			&	20210521	&	F736		&	42-43	&	100	&	90$\times$90	&	78	&	8(3)	&	2496	\\
		&			&		&				&		&				&		&	&	Total time	& 5928 (1.65h)	\\
		& D (153 \um)	&	20210520	&	F735		&	41-43	&	200	&	100$\times$100	&	78	&	18	&	5616	\\
		&			&	20210521	&	F736		&	42	&	200	&	100$\times$100	&	78	&	5	&	1560	\\
		&			&		&				&		&				&		&	&	Total time	& 7176 (1.99h)	\\
		& E (214 \um)	&	20210506	&	F727		&	39-40	&	200	&	180$\times$180	&	78	&	3(3)	&	936\\
		&			&			&			&				&		&	210$\times$210	&	78	&	7(1)	&	2184 \\
		&			&	20210507	&	F728		&	39	&	200	&	210$\times$210	&	78	&	0(1)	&	0 \\
		&			&	20210511	&	F729		&	39	&	200	&	210$\times$210	&	78	&	2	&	624 \\
		&			&	20210513	&	F731		&	43	&	200	&	180$\times$180	&	78	&	1(2)	&	312	\\
		&			&		&				&		&				&		&	&	Total time	& 4056 (1.13h)	\\
\hline
M~83	& D (154 \um)	&	20210512	&	F730	&	38-41	&	200	&	100$\times$100	& 120	&	15 (1) 	& 6720 \\
		& 			&	20210513	&	F731	&	41		&	200	&	100$\times$100	& 120	&	 8	&	3840	\\
		&   			&	20210520	&	F735	&	39-41	&	200	&	100$\times$100	& 120, 60	& 	16, 1	& 7920 \\
		&  			&	20210521	&	F736	&	39-42	&	200	&	 100$\times$100	& 120	&	10	&	4800	\\
		&			&			&		&				&		&					&		&	Total time	& 23760 (6.60h)	\\
\hline
NGC~253	& C (89 \um)	&	20210901	&	F776		&	43	&	100	&	70$\times$70	&	100	&	9	&	3600	 \\
		&			&		&				&		&				&		&	&	Total time	& 3600 (1.00h)	\\
		& D (154 \um)	&	20210915	&	F784		&	43	&	100	&	90$\times$90	&	100	&	7(1)	&	3200	\\
		&			&		&				&		&				&		&	&	Total time	& 3200 (0.89h)	\\
\hline
NGC~1068	&	A (53 \um)		&	20210828	&	F774	&	43	&	100	&	60$\times$60	&	120	&	3	&	1440 \\
			&				&	20211103	&	F786	&	43	&	100	&	60$\times$60	&	120, 45	&	6(1), 1 &	3060 \\
			&				&			&		&				&		&				&	&	Total time	& 4500 (1.06h)	\\
			&	C (89 \um)	&	20211207	&	F800	&	40-42	&	100	&	70$\times$70	&	120, 60 	&	11, 1	& 5520 \\
		 	&				&	20211209	&	F802	&	42-43	&	100	&	70$\times$70	&	120, 60	&	3(1), 1	& 1680 \\	
			&				&			&		&				&		&				&	&	Total time	& 7200 (2.00h)	\\
\hline
NGC~1097$^{\star}$	&	D (154 \um)	&	20200128	&	F654		&	38	&	200	&	90$\times$90	&	120	&	2	&	960	\\
		&			&		&				&		&				&		&	&	Total time	& 960 (0.27h)	\\
\hline
NGC~2146	&	A (53 \um)		&	20200125	&	F653		&	41-42	&	100, 60	&	30$\times$30, 60$\times$60	&	100	&	2, 1	&	1200 \\
			&				&			&			&				&	100	&	60$\times$60	&	60	&	2	&	480	\\
			&				&	20200131	&	F657		&	40-42	&	100	&	30$\times$30	&	100	&	3(2)	&	1200	\\
			&				&			&			&				&		&				&	60	&	7	&	1680	\\
			&				&	20210828	&	F774		&	43		&	100	&	30$\times$30	&	100	&	0(12)	&	0 \\
			&				&	20210831	&	F775		&	41		&	100	&	30$\times$30	&	100	&	3	&	1200	\\
			&			&		&				&		&				&		&	&	Total time	& 5760 (1.60h)	\\
			&	C (89 \um)	&	20200130	&	F656		&	41	&	100	&	50$\times$50	&	100	&	5(1)	& 2000	\\
			&				&	20200131	&	F657		&	42	&	100	&	50$\times$50	&	80	&	4(1)	& 1280	\\
			&				&	20210914	&	F783		&	40-43	&	100	&	50$\times$50	&	100	&	10(1)		&	4000	\\
			&			&		&				&		&				&		&	&	Total time	& 7280 (2.02h)	\\
			&	D (154 \um)	&	20200131	&	F657		&	42-43	&	100	&	70$\times$70	&	80	&	2	&	640	\\	
			&				&			&			&				&		&	90$\times$90	&	80, 60	&	4(1), 1	&	1200	\\
			&				&	20210512	&	F730		&	42-43	&	200	&	100$\times$100	&	100	&	12(3)		& 3600	\\
			&				&	20210911	&	F782		&	41-42	&	200	&	100$\times$100	&	100	&	4(3)	& 1600 \\
			&				&	20210914	&	F783		&	40	&	200	&	100$\times$100	&	100	&	4(1)	&	1600 \\
			&			&		&				&		&				&		&	&	Total time	& 8640 (2.40h)	\\
			&	E (214 \um)	&	20210512	&	F730		&	43	&	200	&	120$\times$120	&	100	&	4(2)	&	1600	\\
			&				&	20210519	&	F734		&	39	&	200	&	120$\times$120	&	100	&	2(3)	&	800	\\
			&				&	20210831	&	F775		&	41-43	&	200	&	120$\times$120	&	100	&	15	&	6000	\\		
			&			&		&				&		&				&		&	&	Total time	& 8400 (2.33h)	\\
\hline
NGC~3627	&	D (154 \um)	&	20210511	&	F729		&	40-41	&	200	&	110$\times$110		&	120	& 12(1)	&	5760	\\
			&				&	20210514	&	F732		&	39-41	&	200	&	110$\times$110		&	120	& 11(3)	&	5280	\\
			&				&	20210519	&	F734		&	38	&	200	&	110$\times$110	&	120	&	11(1)		&	5280	\\
			&			&		&				&		&				&		&	&	Total time	& 16320 (4.53h)	\\
\hline
NGC~4736	&	D (154 \um)	&	20211207	&	F800		&	43	&	200	&	110$\times$100	&	102, 90	&	2, 12	&	5136 \\
			&				&	20211208	&	F8001	&	43	&	200	&	170$\times$170	&	102	&	7	&	2856 \\
			&				&			&			&			&		&					&	&	Total time	& 7992 (2.22h)	\\
\hline
NGC~4826	&	C (89 \um)	&	20210520	&	F735		&	43	&	200	&	110$\times$110		&	100	&	4	&	1600	\\
			&				&	20210521	&	F736		&	43	&	200	&	110$\times$110		&	100	&	7	&	2800	\\
			&			&		&				&		&				&		&	&	Total time	& 4400 (1.22h)	\\
\hline
NGC~6946	&	D (154 \um)	&	20200909	&	F683		&	41-42	&	200	&	120$\times$120	&	100	&	1		&	400	\\
			&				&			&			&	42-43	&	200	&	180$\times$180	&	80	&	17	&	5440	\\
			&				&			&			&	43	&	200	&	220$\times$220	&	80	&	1	&	360	\\
			&				&	20200910	&	F684		&	43	&	200	&	120$\times$120	&	80	&	1		&	320	\\
			&				&			&			&			&		&	200$\times$200	&	80, 60	&	2(4), 5(2)		&	1840 	\\
			&				&	20200911	&	F685		&	43	&	200	&	120$\times$120	&	80	&	2		&	640	\\	
			&				&			&			&			&		&	180$\times$180	&	100	&	14		&	5600	\\
			&				&	20200922	&	F688		&	41-43	&	200	&	200$\times$200	&	60, 90	&	12, 3	&	3960	\\
			&				&	20200923	&	F689		&	42-43	&	200	&	180$\times$180	&	80	&	10(1)	&	3200	\\
			&				&	20200924	&	F690		&	41	&	200	&	200$\times$200	&	80	&	13(1)		&	4160	\\
			&			&		&				&		&				&		&	&	Total time	& 25920 (7.20h)	\\	
\hline
NGC~7331	&	D (154 \um)	&	20210511	&	F729		&	43	&	200	&	120$\times$120	&	120	& 9	&	4320	\\
			&				&	20210513	&	F731		&	43	&	200	&	120$\times$120	&	120	& 7	&	3360 \\
			&				&	20210514	&	F732		&	42-43	&	200	&	120$\times$120	&	120, 90	&	12(1), 1	&	6120	\\
			&				&	20210518	&	F733		&	43	&	200	&	120$\times$120	&	120	&	2	&	960	\\
			&				&	20210519	&	F734		&	40-43	&	200	&	120$\times$120	&	120, 60	&	9, 1	&	4560	\\	
			&				&	20210903	&	F778		&	43	&	200	&	120$\times$120	&	120, 60	&	3, 1	&	1680	\\
			&				&	20210910	&	F781		&	43	&	200	&	120$\times$120	&	120	&	2(4)	&	960	\\
			&				&	20211103	&	F786		&	42	&	200	&	120$\times$120	&	120	&	(4)	&	0	\\
			&			&		&				&		&				&		&	&	Total time	& 21960 (6.10h)	\\			
\enddata
\tablenotetext{{\star}}{Circinus observations from program ID 07\_0032 (PI: Lopez-Rodriguez, E.) and NGC~1097 observations from program ID 07\_0034 (PI: Lopez-Rodriguez, E.).}
\end{deluxetable*}

\subsection{Data reduction of the on-the-fly-map (OTFMAP) polarimetric mode}\label{subsec:DR}

\begin{figure*}[ht!]
\includegraphics[angle=0,scale=0.38]{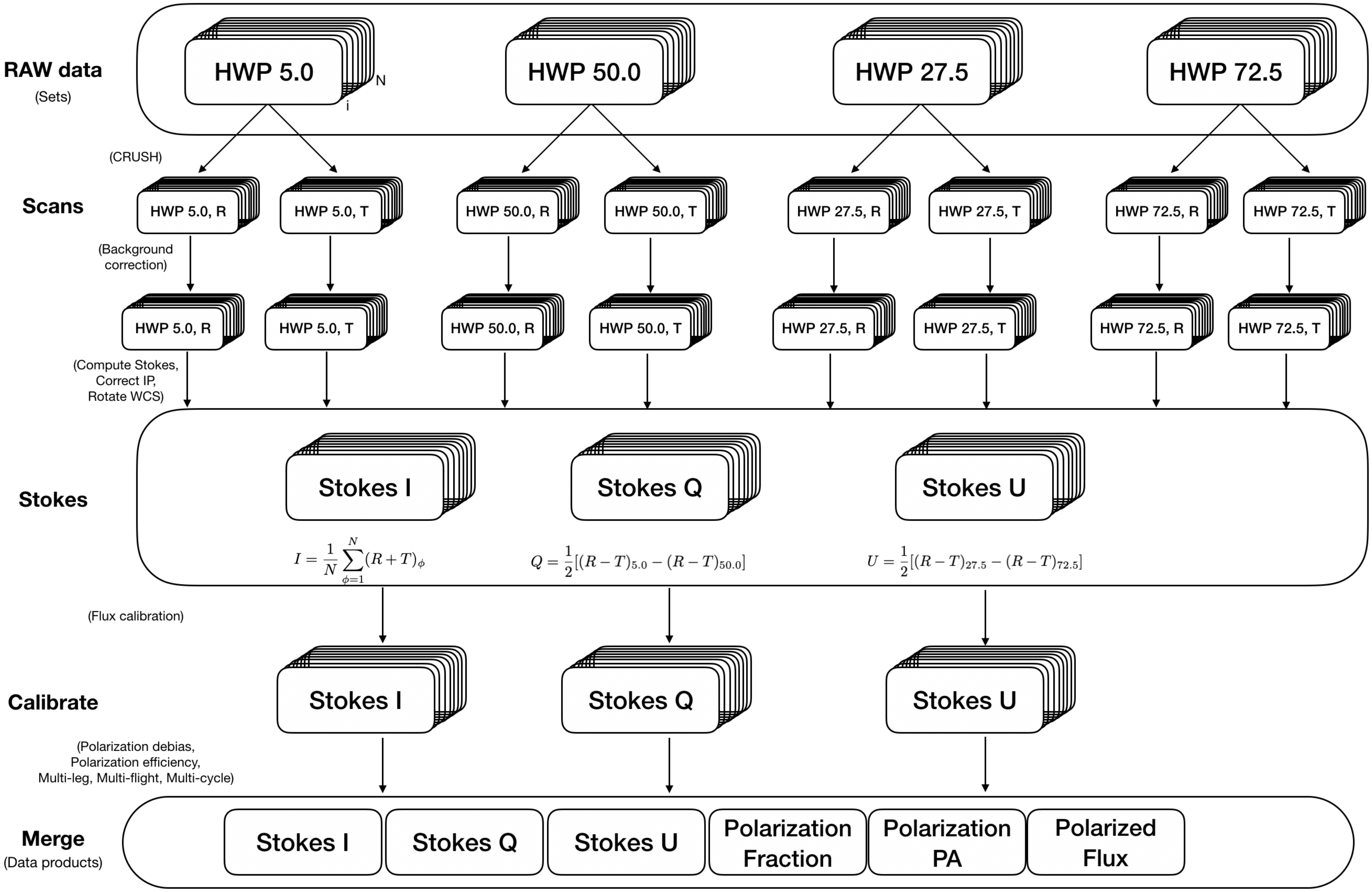}
\caption{On-the-fly-map (OTFMAP) polarimetry data reduction flow-chart. 
The raw data is combined in sets of four half-wave plate position angles (HWP~PA) within a sky rotation tolerance of $3^{\circ}$. The sets are reduced using \textsc{crush}, which generates a scan per $\mathcal{R}$ and $\mathcal{T}$  arrays and per HWP~PA. Then, these scans are zero-level background corrected and combined, from which Stokes parameters are derived. After the Stokes parameters are corrected for instrumental polarization, the scans are sky rotated and flux calibrated. Finally, all Stokes parameters from different legs, flights, and/or cycles are merged to compute the final images, which contain Stokes $IQU$, polarization fraction, $P$, position angle, $PA$, of polarization, and polarized fluxes, $PI$. The pipeline computes uncertainties and covariances, which are taken into account during all stages of the data reduction. 
\label{fig:fig1}}
\epsscale{2.}
\end{figure*}

We performed observations using the OTFMAP polarimetric mode with HAWC+. This technique was an experimental observing mode used during the SOFIA Cycle 7 observations as part of engineering time (IDs: 71\_0019 and 71\_0023, PI: Lopez-Rodriguez, E.) to optimize the polarimetric observations of HAWC+. This observing mode has been routinely used during SOFIA Cycles 8 and 9 for further characterization as part of the programmatic responsibilities of this SOFIA Legacy Program and offered in `shared-risk' mode for those proposals that desired to test this new observing mode. This technique was successfully applied to both galaxies (i.e. Centaurus~A \citep{ELR2021} and NGC~1097 \citep{ELR2021c}) and  Galactic star-forming regions (the filamentary cloud Taurus/L1495 \citep{Li2021}, and OMC-3/4; Li et al., submitted to ApJ). These early results were part of the SOFIA Cycle 7 and 8 observations. Here, we describe the details of the OTFMAP polarimetric mode as the new observing strategy for HAWC+ on the study of magnetic fields in galaxies.

\subsubsection{Processing raw data using \textsc{crush}}\label{subsec:crush}

We reduced the raw data using the Comprehensive Reduction Utility for SHARP II v.2.50-1 \citep[\textsc{crush};][]{kovacs2006,kovacs2008} and the \textsc{hawc\_drp\_v2.7.0} pipeline developed by the data reduction pipeline group at the SOFIA Science Center\footnote{A full description of the most recent pipeline by the SOFIA Science center can be found at \url{https://www.sofia.usra.edu/data/data-analysis}}. SOFIA provided a version, \textsc{hawc\_drp\_v3.0.0}, that replaces \textsc{crush}, written in \textsc{java}, with a new version written in \textsc{python}. This new version\footnote{SOFIA data pipeline in \textsc{python} can be found at: \url{https://github.com/SOFIA-USRA/sofia_redux} and \url{https://www.sofia.usra.edu/data/data-pipelines}} is a one-for-one drop-in replacement with all \textsc{crush}'s options included. The analysis presented here still holds for this new version.

HAWC+ polarimetric observations simultaneously measure two orthogonal components of linear polarization arranged in two arrays of $32 \times 40$ pixels each, labeled as reflective ($\mathcal{R}$) and transmissive ($\mathcal{T}$). Table \ref{tab:HAWC} shows the characteristics of the HAWC+ bands, pixel scales, beam sizes (FWHM), and FOVs of the array for the polarimetric observations. For more information about HAWC+, we refer the reader to \citet{Harper2018}.

\centerwidetable
\begin{deluxetable}{cccccc}
\tablecaption{HAWC+ configuration. \emph{Columns, from left to right:} a) Band name. b) Central wavelength of the band in \um. c) FWHM bandwidth in \um. d) Pixel scale of the band in \arcsec. e) FWHM in \arcsec. and f) FOV for polarimetric observations in \arcmin.
\label{tab:HAWC} 
}
\tablecolumns{6}
\tablewidth{0pt}
\tablehead{\colhead{Band}	&	\colhead{$\lambda_{c}$} &	\colhead{$\Delta \lambda$}	&	\colhead{Pixel scale}	&	\colhead{Beam size}	& \colhead{Polarimetry FOV}  \\ 
 	& \colhead{(\um)}	& \colhead{(\um)} &  \colhead{ (\arcsec)} & \colhead{(\arcsec)}  & \colhead{(\arcmin)} \\
\colhead{(a)} & \colhead{(b)} & \colhead{(c)} & \colhead{(d)} & \colhead{(e)} & \colhead{(f)} 
}
\startdata
A	& $53$	&	$8.7$&	$2.55$	&	$4.85$	&	$1.4\times1.7$ \\
C	& $89$	&	$17$	&	$4.02$	&	$7.80$	&	$2.1\times2.7$	\\
D	& $154$	&	$34$	&	$6.90$	&	$13.6$	&	$3.7\times4.6$	\\
E	& $214$	&	$44$	&	$9.37$ 	&	$18.2$	&	$4.2\times6.2$  \\	
\enddata
\end{deluxetable}

The OTFMAP polarimetric mode performs observations in a sequence of four Lissajous scans, where each scan has a different halfwave plate (HWP) position angle (PA) in the following sequence: $5^{\circ}$, $50^{\circ}$, $27.5^{\circ}$, and $72.5^{\circ}$. A starting angle of $5^{\circ}$ is chosen to avoid the reference position, $0^{\circ}$, of the HWP \citep{Harper2018}.  This sequence is called a `set' hereafter (Figure \ref{fig:fig1}-first row). In this HAWC+ observing mode, the telescope is driven to follow a parametric curve with a nonrepeating period whose shape is characterized by the relative phases and frequency of the motion. Each scan is characterized by its  amplitude,  rate,  angles, and  duration.  Table \ref{tab:OBS} shows the Lissajous parameters for each galaxy (examples of Lissajous patterns are shown in Figures \ref{fig:fig9} and \ref{fig:figA3}). 

Each set was reduced using \textsc{crush}, which estimates and removes the correlated atmospheric and instrumental signals, solves for the relative detector gains, and determines the noise weighting of the time streams in an iterative pipeline scheme. Each reduced scan produces two images associated with each array, $\mathcal{R}$ and $\mathcal{T}$. Both images are orthogonal components of linear polarization at a given HWP~PA. At this stage, eight different scans are computed (Figure \ref{fig:fig1}-second row).

One of the limitations of the OTFMAP observing mode lies in the recovery of large-scale diffuse and faint emission from the astrophysical objects. This challenge is a result of the finite size of the array, variable atmosphere conditions, variable detector temperatures, scan rates, and the applied filters in the reduction steps used to recover extended emission. We applied several filters using \textsc{crush} to recover large-scale emission structures of the galaxies while paying close attention to any change that may compromise the intrinsic polarization pattern of the astrophysical source. A total of 16 different pipeline schemes (e.g. \textsc{faint, extended, bright}, number of iterations, several filtering options) were performed per galaxy per band. For each of these pipeline schemes, the full data reduction steps were performed and statistical comparisons between the Stokes parameters $IQU$ were computed. Specifically, flux conservation was compared between pipeline schemes. The morphologies between Stokes $IQU$ per pipeline scheme and between Stokes $I$ and PACS/\textit{Herschel} observations at the closest wavelength were cross-correlated, and Anderson-Darling tests were conducted using the polarization fraction and angles between pipeline schemes. In addition, final products using these different schemes were compared to the galaxies with available C2N observations (M82, NGC~1068, and Circinus). We find that any additional options to the pipeline configuration described above introduce artificial polarization patterns to the galaxies. Some examples of the artificial polarization patterns include:  a) constant PA of polarization across the FOV, b) changes of the PA of polarization larger than $30^{\circ}$ in regions of high signal-to-noise ratio in Stokes $I$ (SNR$_{I}\ge500$), c) additional extended structures in Stokes $I$ that were not present in \textit{Herschel} images, d) flux calibration incompatible with \textit{Herschel} images, and e) differences in the PA of polarization of up to $50^{\circ}$ when compared with the C2N observing mode. Thus, the standard pipeline using the nominal configuration for \textsc{crush} provided by the SOFIA Science Center was used. 

The angular extension of all galaxies are within the polarimetry FOVs of the array in any given band (Table \ref{tab:HAWC}). For all observations, the final FOV of the images have zero-level background which enables a measure of the true sky background without contribution from the astrophysical object. Thus, the standard pipeline recovers the large-scale extended emission of the galaxies in our sample. In contrast, observations of molecular clouds or the Galactic center may have extended  emission much larger than the polarimetry FOV in any given band. We point the reader to Taurus/L1495 observed with HAWC+ at 214 \um, where further pipeline steps were required to recover those large-scales \citep{Li2021}. For all the galaxies in our sample, observations with a sky rotation smaller than $3^{\circ}$ were reduced within the same set, which increases the signal-to-noise ratio (SNR) per pixel to optimize the correlation between the astrophysical, sky, and instrumental signals. This approach optimizes the recovery of large-scale and low-surface brightness of the science object while keeping the rotation of the polarization angle vector within the intrinsic angular uncertainty of $3^{\circ}$ of HAWC+ \citep{Harper2018}.

Table \ref{tab:OBS} shows several rejected sets for some galaxies, where $13$\% of the data was removed from the full sample. In general, these sets were removed due to tracking issues during the observations, producing WCS offsets in the timestreams of the detector pixels that affect the map-making process. The worst case scenario was found during the observations of NGC~2146 at $89$ \um, flight F774. The object was found to have an offset from the position of the boresight of the array for each HWP~PA during the full length of observations. Post-correction of the WCS as a function of the HWP~PAs: a) did not fully correct the angular offsets, and b) produced a final product with polarization maps incompatible with the other flights. Given these issues, NGC~2146 observations in Band A during flight F774 were removed from the final data products.

\subsection{Zero-level background correction}\label{subsec:ZL}

HAWC+ measures the power of the emissive and variable atmosphere and the astrophysical object. The data reduction scheme described above may produce regions of negative flux in areas of extended and low surface brightness due to the similar levels of noise and astrophysical signal. Thus, characterizing and estimating the zero-level background across the FOV of the observations of galaxies is imperative in mitigating the need to approximating and adding the lost flux back to the full image later. 

We have determined and corrected the zero-level background of our observations as follows. Since all galaxies are smaller than the HAWC+ FOV for a given band and they are isolated objects without large-scale extended thermal emission, the scan amplitudes (Table \ref{tab:OBS}) were selected to have at least 1/3 of the final FOV with `true’ zero-level background counts. In addition, the polarization skydips (Section \ref{subsec:IP}) show that the sky is unpolarized. Thus, the sky background around the galaxies represents the `true’ zero-level, where flux and polarimetric calibration can be performed. Specifically, pixels with a SNR in total intensity $\ge3$  were masked for each of the scans per HWP~PA and $\mathcal{R}$ and $\mathcal{T}$ arrays produced by \textsc{crush} (Figure \ref{fig:fig1}-second row). The masked background was fitted with a second-order surface and then added to the unmasked image, which ensures a positive and flat background across the full FOV. The second-order surface has six free parameters, $f(x,y) = Ax^{2}  + By^{2} + Cxy + Dx + Ey + F$, that were fit to the Nyquist-sampled images using several hundreds pixels. Using the Nyquist sampling or an individual pixel per beam did not affect the final fitting. A second-order surface provides better results than a first-order (i.e. flat) surface. The background has similar curvature as the exposure maps, which have an exposure time that varies with the distance from the center of the image. This radial variation and the weighting computed by \textsc{crush} per iteration produce the observed curvature in the background. As a test, this approach was also applied to unpolarized objects (i.e. planets), and the measured fluxes and instrumental polarization were found to be compatible with the C2N and skydips observing modes (Section \ref{subsec:IP}). We estimated that the total flux from the second-order surface is $\le1\sigma$ of the pixel-to-pixel variation within each of individual scans, and the  polarization associated with it is lower ($\le0.3\%$) than the instrumental polarization ($\sim1.6-2.1$\%, Section \ref{subsec:IP}) of HAWC+ in any given band.

A similar approach was used in Taurus/L1495 by \citet{Li2021}, where a simpler approach to estimating the background in a C2N observation was performed. For these observations, an adjacent region equal to the FOV of the HAWC+ array close to the molecular cloud was identified using \textit{Herschel} images. Using the HAWC+ observations, the mean of the background was estimated and considered as the zero-level background. Finally, the mean was added to the full FOV of the observations. These authors found that this technique contributed $\sim14$\% to the polarized flux in their science observations. Here, we use a second-order surface to fit the masked image after the removal of the astrophysical object, yields lower artificial polarized flux and an optimal correction of the background structure. At this stage, eight different scans with positive and flat background are computed (Figure \ref{fig:fig1}-third row).

\subsection{Precipitable water vapour correction}\label{subsec:PWV}

The standard pipeline corrects for the precipitable water vapour (PWV) based on the altitude of the aircraft at the time of the observations. As real-time PWV data are not available, this correction is performed using a model developed by the SOFIA Science Center that provides flux calibrations within a maximum of $20$\% uncertainty. If PWV variability and/or bad weather conditions were present during observations,  further correction may be required to minimize flux variations that can affect the final flux calibration and/or polarization measurements. 

\begin{figure*}[ht!]
\includegraphics[angle=0,scale=0.35]{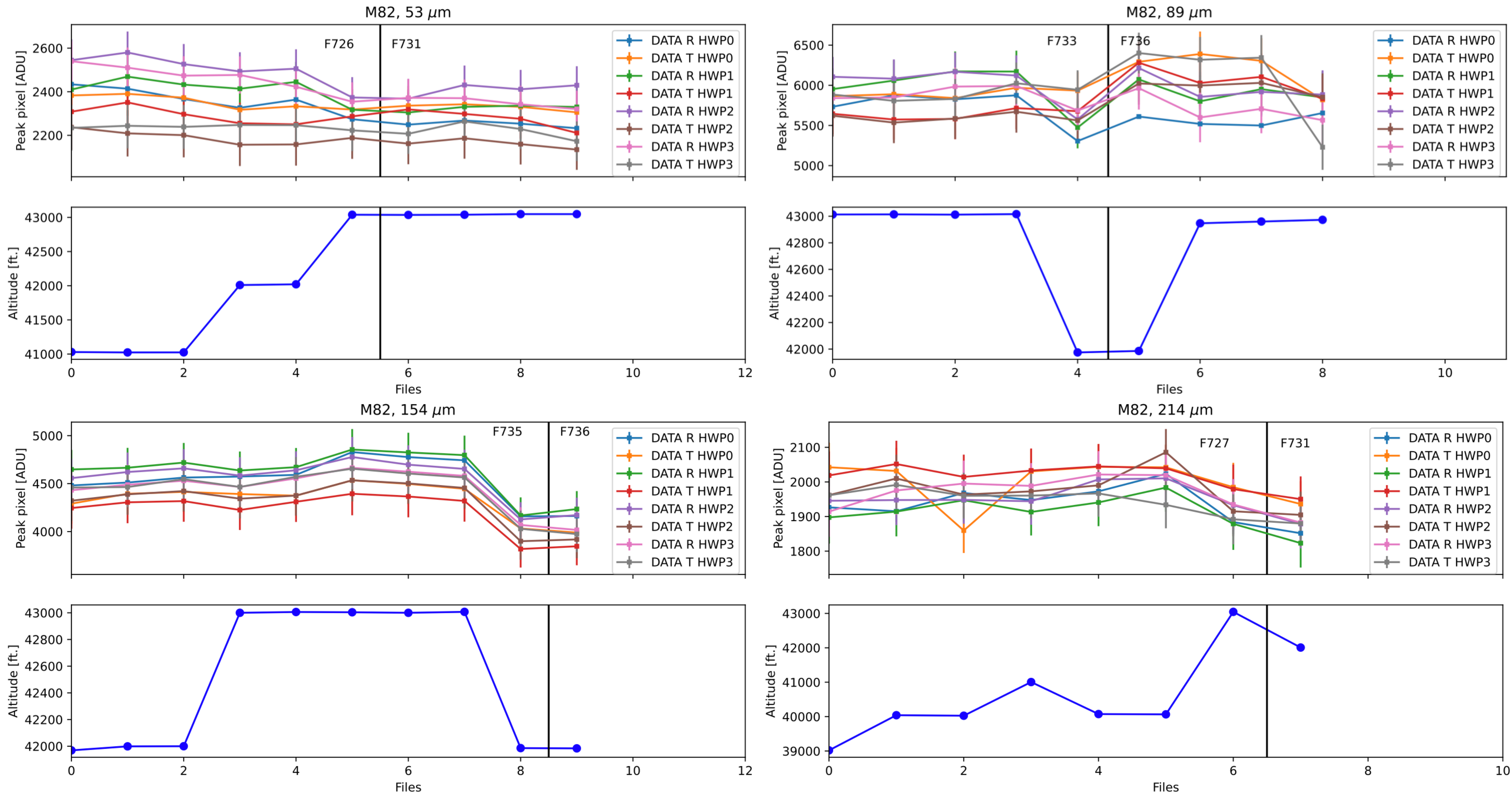}
\caption{Example of flux variation as a function of time and aircraft altitude for M82 at all HAWC+ bands. Peak fluxes (first and third row) of the scans per HWP~PA and $\mathcal{R}$ and $\mathcal{T}$ arrays at 53 (top left), 89 (top right), 154 (bottom left) and 214 (bottom right) \um~of M82. The aircraft altitude (second and fourth rows) at the time of data acquisition is shown. the vertical lines show the boundaries between different flights.
\label{fig:fig2}}
\epsscale{2.}
\end{figure*}

We examined the flux variations of the scans per HWP~PA and array after background correction for all galaxies as a function of the altitude and weather conditions. All galaxies show observed flux variations $\le15\%$ at all bands independently of the altitude and weather conditions. Figure \ref{fig:fig2} shows an example of flux variations as a function of time and aircraft altitude for all HAWC+ bands for M82. This figure demonstrates that flux variations are negligible during climbing in the flight profile. Also, the relative flux contributions from HWP~PAs and arrays are mostly constant, although the relative mean flux of the full scan may vary within the maximum 20\% flux uncertainty. This result implies that the intrinsic polarization of the source is conserved relative to the altitude and weather conditions within a set of four HWP~PAs.

\begin{figure*}[ht!]
\includegraphics[angle=0,scale=0.45]{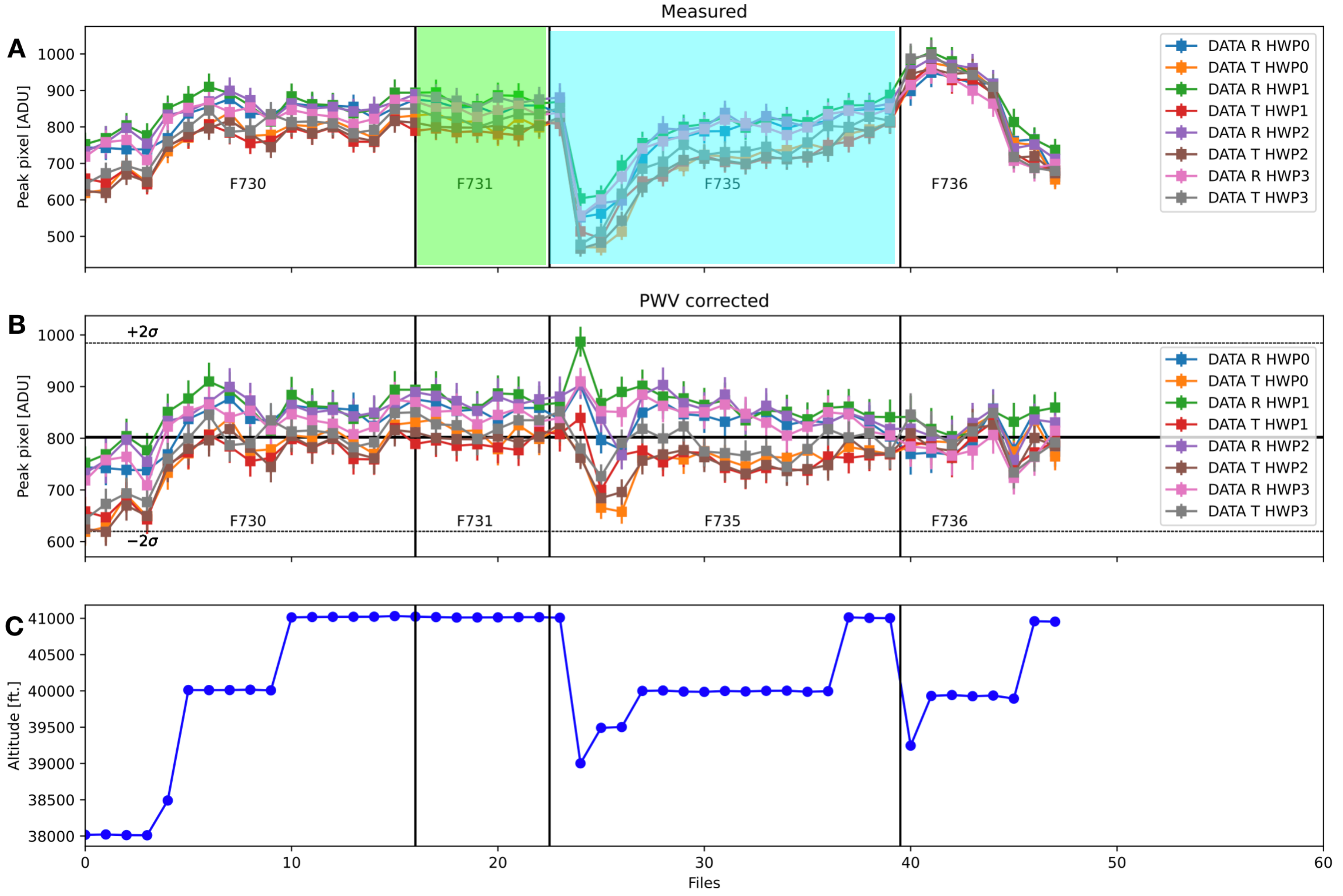}
\caption{Flux variation and aircraft altitude as a function of time for the observations of M83 at 154 \um. A) Measured peak fluxes of the scans per HWP~PA and $\mathcal{R}$ and $\mathcal{T}$ arrays; B) PWV corrected fluxes using the fits shown in Figure \ref{fig:fig4}. The mean (black solid line) and $2\sigma$ uncertainty (black dotted line) are shown. C) Aircraft altitude at the time of data acquisition. The PWV during the observations of M83 (green for F731 and cyan for F735 shadowed regions) is shown in Figure \ref{fig:fig3b}.
\label{fig:fig3a}}
\epsscale{2.}
\end{figure*}

\begin{figure*}[ht!]
\centering
\includegraphics[angle=0,scale=0.4]{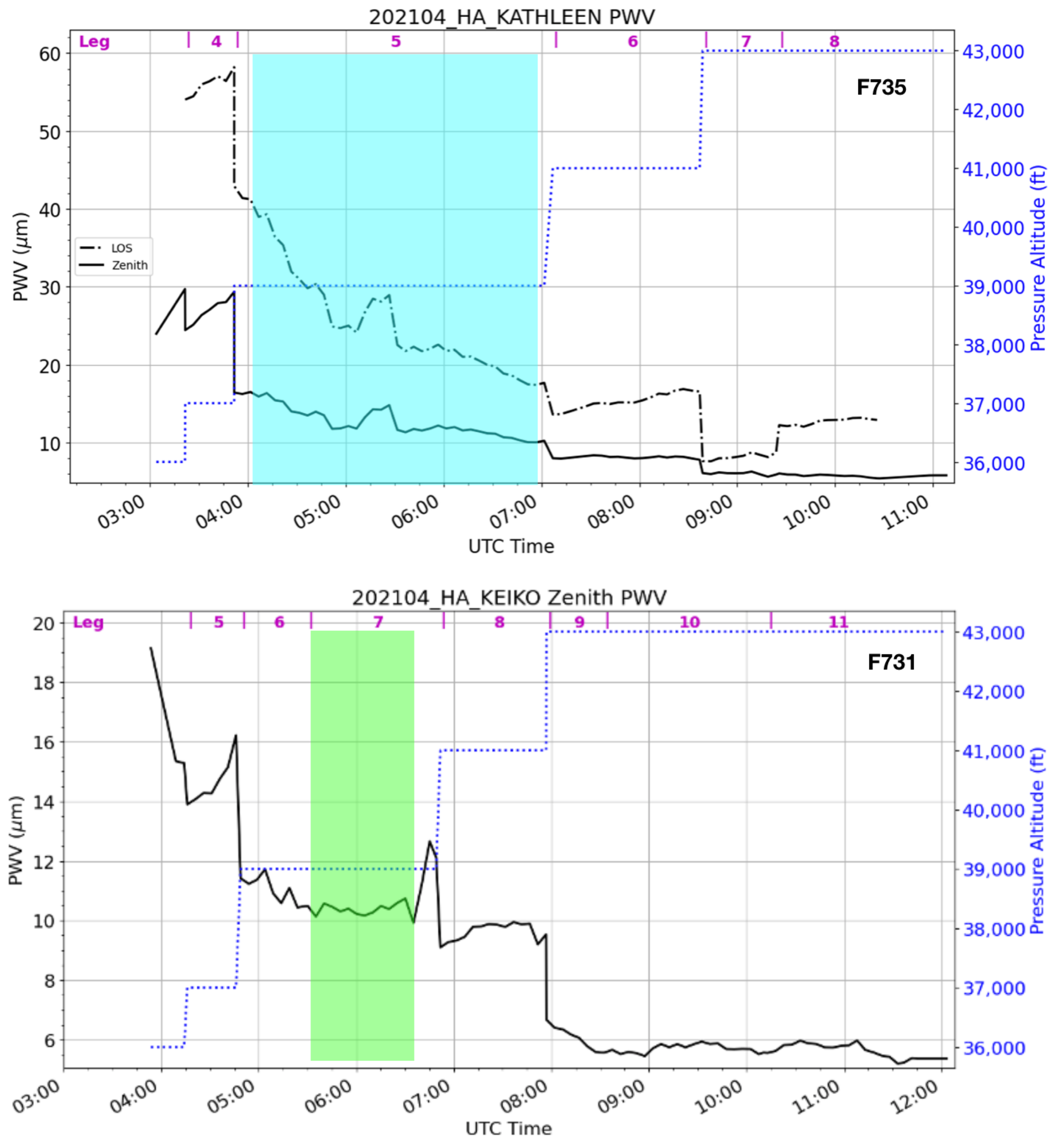}
\caption{PWV variation as a function of time for the flights associated with M83 at 154 \um~(see Figure \ref{fig:fig3a}).  Top panel: expected PWV for flight F735. The PWV during the observations of M83 (cyan shadowed region) is shown. Bottom panel: expected PWV for flight F731. The PWV during the observations of M83 (green shadowed regions) is shown. 
\label{fig:fig3b}}
\epsscale{2.}
\end{figure*}

\begin{figure}[ht!]
\centering
\includegraphics[angle=0,scale=0.60]{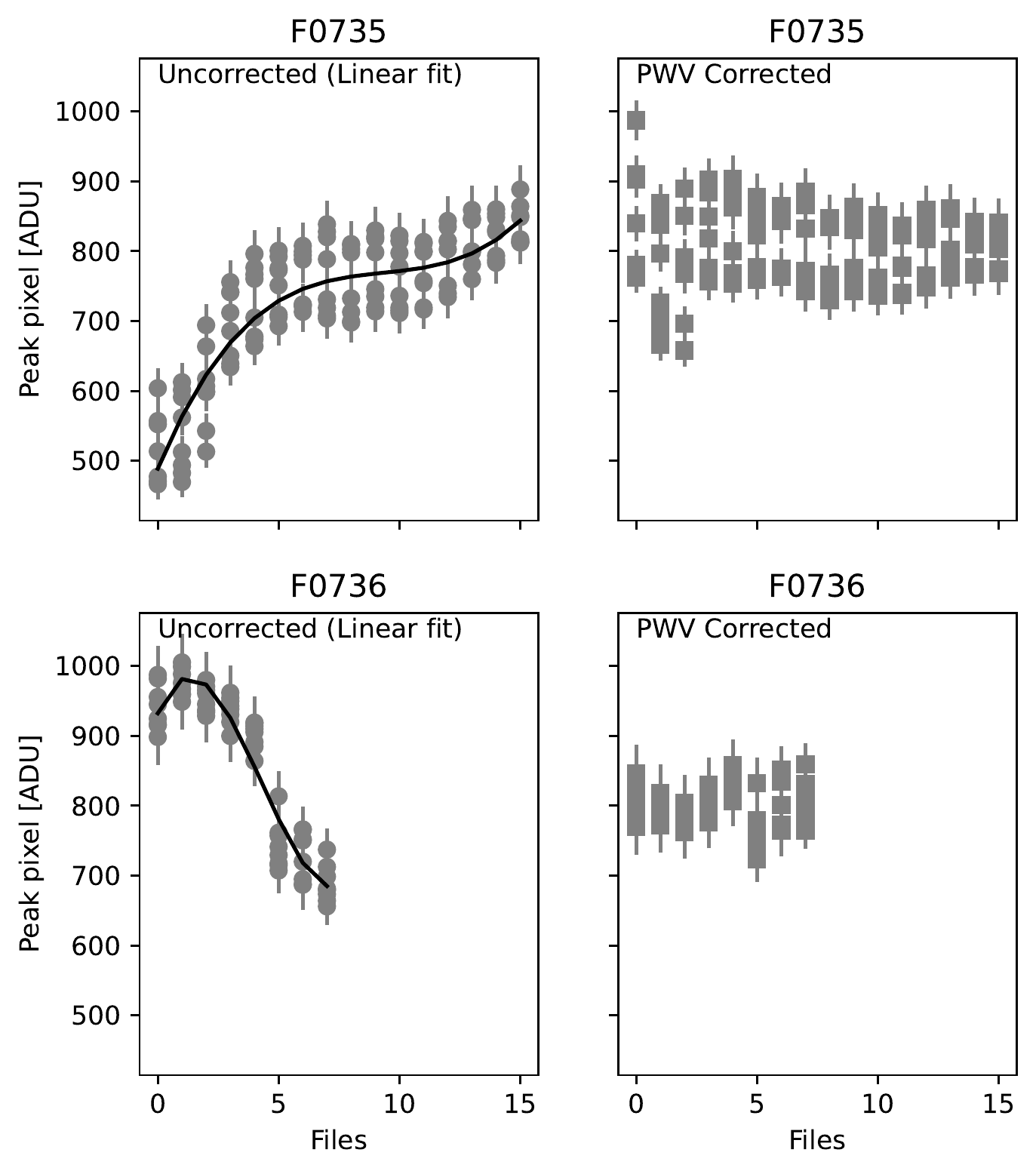}
\caption{Fluxes uncorrected and corrected due to PWV variations during observations of M83. Measured fluxes (left column) are fit with a polynomial function of order 3 (black solid line) for flights F0735 (top), F0736 (bottom). Corrected fluxes (left column) by the normalized fit to the mean flux of flight F731 are shown.
\label{fig:fig4}}
\epsscale{2.}
\end{figure}

The only exceptions are M83 and NGC~6946. We found that these galaxies have strong flux variations during several days of observations that cannot be explained by altitude changes or instrumental performance. Figure \ref{fig:fig3a} shows the measured flux variations for M83 at 154 \um. Flights F735 and F736 have flux variations that exceed the nominal $20$\% flux uncertainties provided by HAWC+. These flux variations are not directly related to changes in the aircraft altitude (Figure \ref{fig:fig3a}-A vs. \ref{fig:fig3b}-Bottom). Instead, changes in the PWV during the data acquisition explain the flux variations. Specifically, Flight F731 has a constant mean flux across the 1.5h of observations with also a constant PWV of $\sim11$ \um. However, the flux increases by a factor of 1.8 in flight F735, while the altitude is constant across two thirds of the observations with a climb of 2000ft. and 1000ft. at the beginning and end of the 2.5h of observations. The expected PWV for flight F735 also decreases by a factor 1.8 during the observations (Figure \ref{fig:fig3b}-Top) and shows a bump at approximately halftime of the observations. This behavior causes an increase in flux as well as a bump in the flux measured by HAWC+, which is in agreement with the measured fluxes and the factor of the increased fluxes (Figure \ref{fig:fig3a}-A vs. \ref{fig:fig3b}-Top, as well as Figure \ref{fig:fig4}). 

We correct the PWV variations for each flight as follows. Fluxes for flights F735 and F736 were fitted with a polynomial function of order 3 and rescaled to the mean flux from flight F731. Figure \ref{fig:fig4}-left shows the measured (uncorrected) fluxes with the fitted polynomial function. To conserve the relative contribution of the fluxes between $\mathcal{R}$ and $\mathcal{T}$ scans for each HWP PA per set, the polynomial function was fit using all measured fluxes as shown in Figure \ref{fig:fig4}. The fit was normalized to the mean measured flux from F731, which shows a constant flux within the flight at the highest altitude from the full set of observations. Finally, each measured flux from each set was multiplied by the normalization factor to correct for the PWV variation (Figure \ref{fig:fig4}-right). Final corrected fluxes are shown in Figure \ref{fig:fig3a}-B. In average, fluxes vary by $\sim15$\% across the full observations with some outlier sets at $\sim20$\%. The same methodology was applied to NGC~6946 (Appendix \ref{App:PWV}).  At this stage, eight different scans with positive background and corrected by PWV varations are computed (Figure \ref{fig:fig1}-third row).

\subsection{Stokes parameters and instrumental polarization}\label{subsec:IP}

\begin{figure}[ht!]
\centering
\includegraphics[angle=0,scale=0.60]{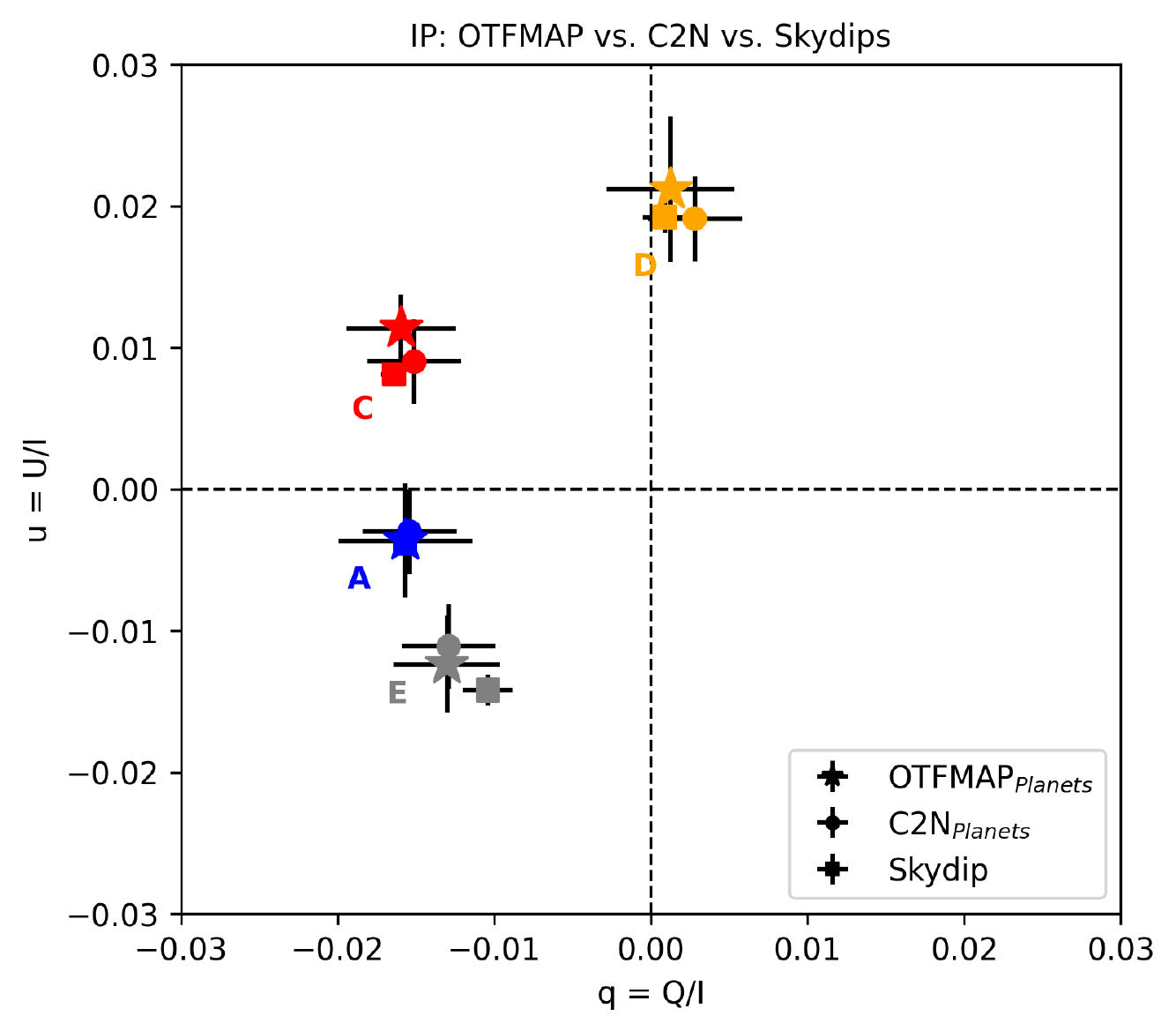}
\caption{Instrumental polarization for the C2N and OTFMAP observing modes at all bands. IP for the C2N (circles), OTFMAP (stars), and polarization skydips (squares) for bands A (blue), C (red), D (orange), and E (grey) are shown. Definitions and details of the estimation of the IP are described in Section \ref{subsec:IP} and \ref{subsec:IP2}.
\label{fig:fig5}}
\epsscale{2.}
\end{figure}

The Stokes parameters $IQU$ were estimated using the double difference method (equations shown in Figure \ref{fig:fig1}) in the same manner as the standard C2N observations carried by HAWC+ described in Section 3.2 by \citet{Harper2018}. In Fig.\ref{fig:fig1}, $N$ is the number of sets, and $\mathcal{R}_{\phi}$ and $\mathcal{T}_{\phi}$ are the total intensity images of the $\mathcal{R}$ and $\mathcal{T}$ arrays per HWP~PA. The instrumental polarization (IP) was corrected using the polarization skydips. Specifically, the polarization skydips were performed by moving the telescope from $57^{\circ}$ to $23^{\circ}$ in elevation while the HWP rotates at a constant speed \citep{Harper2018}. These observations provide an estimate of the Stokes $qu$ per detector pixel across the full array (Appendix \ref{App:IP}, Figure \ref{fig:figA3n}). The final IP was estimated as the weighted median of all pixels and the associated uncertainty was estimated as the weighted standard deviation of the median. Figure \ref{fig:fig5} shows the estimated IP as a percentage of the normalized Stokes $qu$ in the instrument reference frame. The numerical values and their uncertainties are shown in Table \ref{tab:IP}. We estimate an IP variation of $\le0.2$\% across the FOV with a median IP of $1.6-2.1$\% within the wavelength range of $53-214$ \um.

For all galaxies, the IP was corrected by subtraction a constant Stokes $qu$ from the measured Stokes $qu$ of each galaxy. We use the normalized Stokes $qu$ from the polarization skydips. The IP correction was performed in the instrument reference frame before the rotation to the sky coordinates. This correction has a polarization uncertainty of $\le0.2$\% across the FOV in any band \citep[][and Section \ref{subsec:IP2}, Figure \ref{fig:figA3n}]{Harper2018}. We include an uncertainty of $0.2$\% in the polarization fraction in quadrature to the uncertainties estimated in the scientific analysis of this project. Finally, for each set, we rotate the Stokes $QU$ from the instrument to the sky coordinates. At this stage, the Stokes $QU$ corrected by IP are in sky coordinates for each set (Figure \ref{fig:fig1}-fourth row). 

To quantify the feasibility of our data reduction scheme, we performed OTFMAP polarimetric observations of planets (i.e. Neptune and Uranus) in all bands during 2021. We reduced these observations using the same approach as described above. Specifically, for each set of four HWP~PAs, we reduced the raw data before estimating the Stokes $IQU$ images. Then, Stokes $QU$ were normalized such that $q=Q/I$ and  $u=U/I$. We computed the median and standard deviation of $q$ and $u$ within the beam size centered at the peak intensity of Stokes $I$ for each band. The final IP was estimated as the weighted median of all measurements and the associated uncertainty was estimated as the weighted standard deviation of the median. As an example, Appendix \ref{App:IP} (Figure \ref{fig:figA3}) shows the Stokes $I$ and normalized Stokes $qu$ for Neptune at $53$ \um. The central beam (white circle) used to estimate the Stokes parameters is shown. After we apply the same IP correction to the planets, we estimate a residual polarization of $\le0.3$\%.

\begin{deluxetable*}{ccccccc}
\tablecaption{Instrumental polarization from OTFMAP, C2N, and skydips. \emph{Columns, from left to right:} a) Band name, b) percentage Stokes $qu$ for the OTFMAP observing mode, c) percentage Stokes $qu$ for the C2N observing mode, and  d) percentage Stokes $qu$ for polarization skydips.
\label{tab:IP} 
}
\tablecolumns{6}
\tablewidth{0pt}
\tablehead{\colhead{Band}	&	\multicolumn{2}{c}{OTFMAP}  &	\multicolumn{2}{c}{C2N}	&	\multicolumn{2}{c}{Skydips}	\\ 
 	&	\colhead{$q$ (\%)}	&	\colhead{$u$ (\%)}	& \colhead{$q$ (\%)}	&	\colhead{$u$ (\%)} &  \colhead{q (\%)}	&	\colhead{u (\%)}\\
\colhead{(a)} &	\colhead{(b)}  & \colhead{(c)} & \colhead{(d)}	&	\colhead{(e)} & \colhead{(f)} & \colhead{{g}}
}
\startdata
    &  \multicolumn{2}{c}{2021$^{\dagger}$}                     & \multicolumn{2}{c}{2017}
                    & \multicolumn{2}{c}{Oct-Nov 2017}        \\
A	&	$-1.57\pm0.43$		&	$-0.36\pm0.40$	 	&	$-1.54\pm0.30$		&	$-0.30\pm0.30$		&	$-1.60\pm0.07$		&	$-0.38\pm0.02$\\
C	&	$-1.59\pm0.35$		&	\phs$1.14\pm0.24$		&	$-1.51\pm0.30$		&	\phs$0.90\pm0.30$		&	$-1.64\pm0.08$		&	\phs$0.82\pm0.07$\\
D	&	\phs$0.13\pm0.41$		&	\phs$2.12\pm0.52$		&	\phs$0.28\pm0.30$		&	\phs$1.91\pm0.30$		&	\phs$0.14\pm0.14$		&	\phs$1.89\pm0.11$\\
E	&	$-1.30\pm0.34$ 	&	$-1.24\pm0.34$		&	$-1.29\pm0.30$		&	$-1.11\pm0.30$		&	$-1.09\pm0.16$		&	$-1.41\pm0.11$\\	
\enddata
\tablenotetext{{\dagger}}{The year in which the data were obtained, with polarization measured using one of the three listed techniques.}
\end{deluxetable*}

\subsection{Flux calibration and data merging}\label{subsec:CAL}

The flux is calibrated using observations of standard sources (i.e. planets, asteroids). The SOFIA Science Center routinely observes standard sources through each observing run and derives the flux calibration factors to be applied to each band. \citet{Harper2018} describes the flux calibration processes in more detail. We used the standard flux calibration factors (associated uncertainties are $\le15$\%) provided by the SOFIA Science Center for each observation. At this stage, Stokes $IQU$ in sky coordinates for each set are flux calibrated (Figure \ref{fig:fig1}-fifth row).

Multi-flight and multi-cycle observations were merged. Stokes $IQU$ were merged using an adaptive weighted average\footnote{A detailed description of the adaptive weighting algorithm used by SOFIA can be found at \url{https://sofia-usra.github.io/sofia_redux/sofia_redux/toolkit/resampling/parameters.html\#adaptive-weighting}} within a common grid with a specified pixel size, where North is up and East is left. The pixelization of this grid (i.e. final data product) is set to match the detector pixel size in a given band (Table \ref{tab:HAWC}), a value equivalent to Nyquist sampling. For each pixel, the weighted average within the FWHM of the associated band is estimated. Thus, the final data product has a Nyquist sampling pixelization with correlated fluxes within the FWHM of the band. The error maps are estimated from the input variances for the pixels involved in each weighted average. 

The polarization fraction, $P$, position angle of polarization, $PA$, and polarized flux, $PI$, are derived from the merged Stokes $IQU$. The polarization fraction is then debiased, $P'=\sqrt{P^{2}-\sigma_{p}^{2}}$, and corrected for polarization efficiency, $P'' = P'/P_{eff}$, where $\sigma_{p}$ is the polarization uncertainty, and $P_{eff}$ is the fractional polarization efficiency of $0.842, 0.939, 0.975, 0.978$ at $53$, $89$, $154$, and $214$ \um, respectively. HAWC+ has an absolute error of $3^{\circ}$ in polarization angle and $0.4$\% in polarization fraction \citep{Harper2018}.

Final Stokes $I$ values were spatially cross-correlated with PACS/\textit{Herschel} images at the closest wavelength to correct the WCS of the HAWC+ images. Small corrections of $1-2$ pixels were required for all HAWC+ observations to match with the  central core of the galaxies observed with \textit{Herschel}. These small offsets were applied to the maps of the Stokes $IQU$, $P$, $PA$, $PI$, and their associated uncertainties. At this stage, Stokes $IQU$, $P$, $PA$, $PI$ values and their associated uncertainties are computed and ready for scientific analysis (Figure \ref{fig:fig1}-sixth row). 

\subsection{Data products}\label{subsec:PMP}

The data products generated within this data release are publicly available on the SOFIA Legacy Program website via \url{http://galmagfields.com/}. The database contains 14 galaxies observed at multiple wavelengths as shown in Table \ref{tab:DR1}. A total on-source time of 47.54h (51.34h including overheads) is presented, which corresponds to 33\% of the total awarded time of 155.70h (including overheads) for this SOFIA Legacy Program. An additional 11.84h (including overheads) of observations for Centaurus~A \citep{ELR2021}, Circinus, M51 \citep{Borlaff2021}, and NGC~1097  \citep{ELR2021c} are included in this data release. Each file contains Stokes $IQU$, $P$, $PA$, $PI$ values and their associated uncertainties with a pixel scale equal to the half-beam size for a given band. The released files have the same file format as table 2 of \citet{Gordon2018}. 

\begin{figure*}[ht!]
\includegraphics[angle=0,scale=0.34]{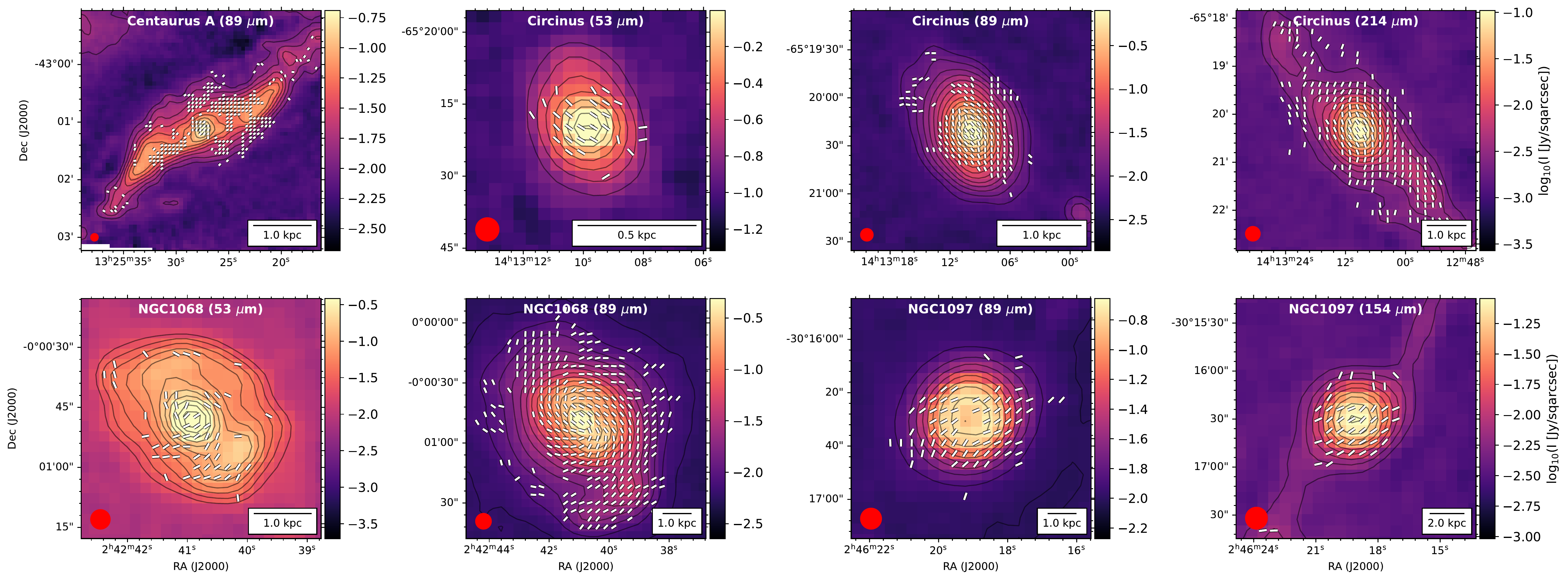}
\caption{Active galactic nuclei. Centaurus~A at 89 \um, Ciricinus at 53, 89, and 214 \um, NGC~1068 at 53 and 89 \um, and NGC~1097 at 89 and 154 \um\ are shown. Total intensity (colorscale) in logarithmic scale with contours increasing in steps of $2^{n}\sigma$, with $n= 5, 5.5, 6, \dots$ The measurements of polarization fraction (white lines) are shown with constant length to display the B-field orientations. Only measurements of polarization fraction with $P/\sigma_{P} \ge 3$ and $I/\sigma_{I} \ge 80$ were selected.  A physical scale (black line) of $0.5$, $1.0$ or $2.0$ kpc and the beam size (red circle) are shown.
\label{fig:fig6}}
\epsscale{2.}
\end{figure*}

\begin{figure*}[ht!]
\includegraphics[angle=0,scale=0.34]{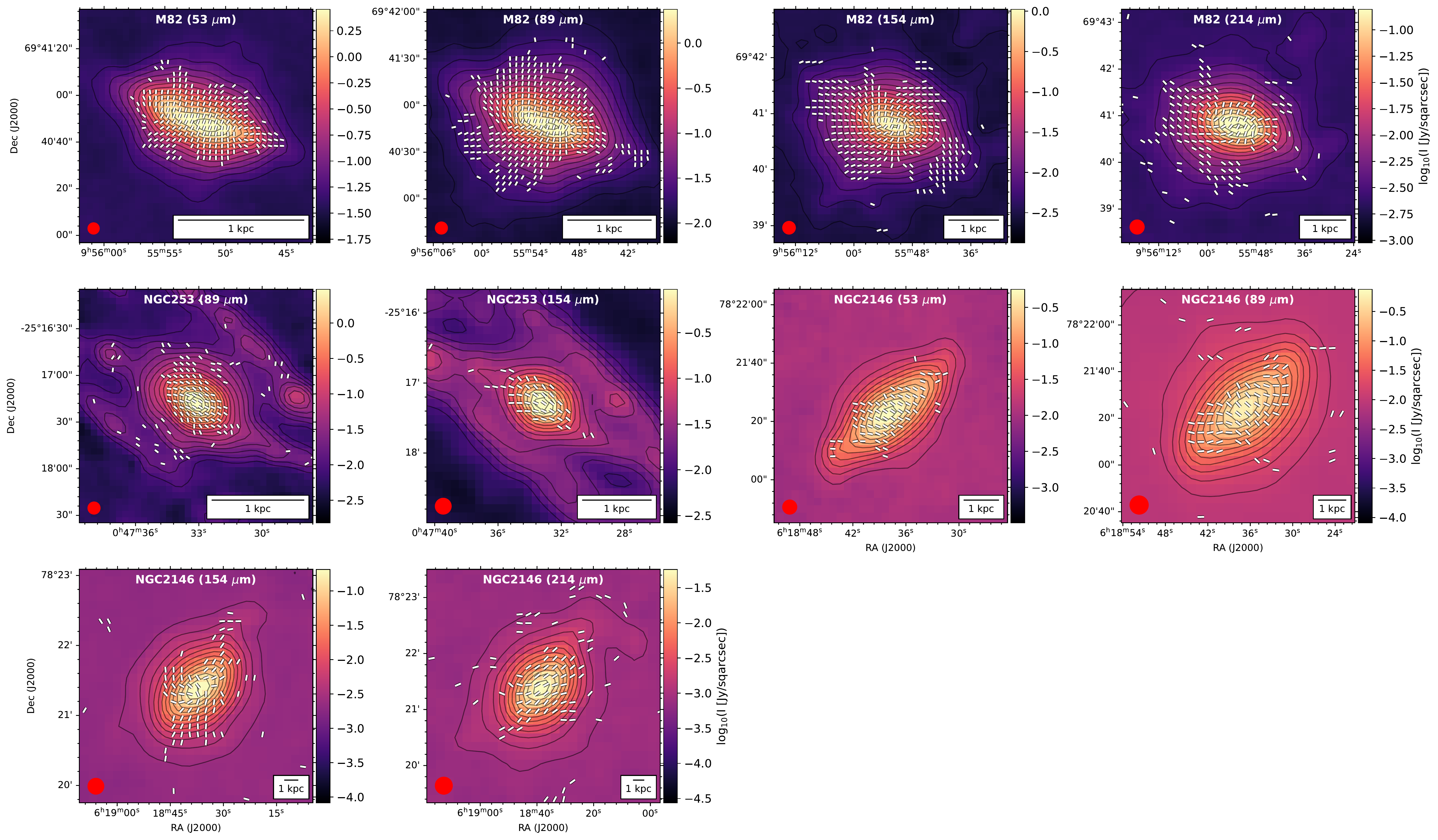}
\caption{Starburst galaxies. M82 at 53, 89, 154, and 214 \um, NGC~2146 at 53, 89, 154, and 214 \um, and NGC~253 at 89 and 154 \um\ are shown. Total intensity (colorscale) is shown in logarithmic scale with contours increasing in steps of $2^{n}\sigma$, with $n= 5, 5.5, 6, \dots$ The measurements of polarization fraction (white lines) are shown with constant length to display the B-field orientations. Only measurements of polarization fraction with $P/\sigma_{P} \ge 3$ and $I/\sigma_{I} \ge 80$ were selected. 
A physical scale (black line) of $1.0$ kpc and the beam size (red circle) are shown.
\label{fig:fig7}}
\epsscale{2.}
\end{figure*}

\begin{figure*}[ht!]
\centering
\includegraphics[angle=0,scale=0.60]{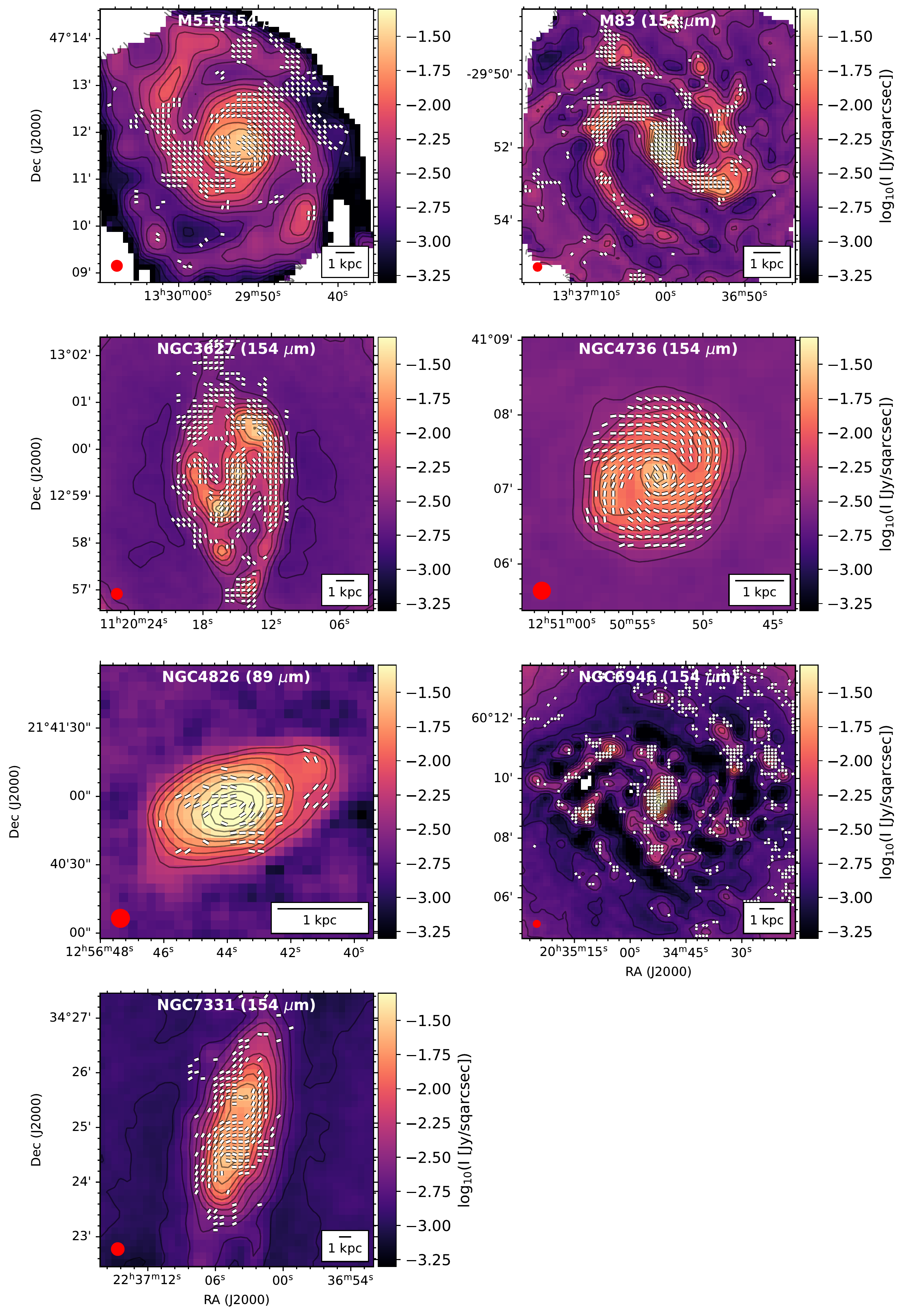}
\caption{Spiral galaxies. M51, M83, NGC~3627, NGC~4736, NGC~6946, and NGC~7331 at 154 \um, and NGC~4826 at 89 \um\ are shown. 
Total intensity (colorscale) is shown in logarithmic scale with contours increasing in steps of $2^{n}\sigma$, with $n= 5, 5.5, 6, \dots$ The measurements of polarization fraction (white lines) are shown with constant length to display the B-field orientations. Only measurements of polarization fraction with $P/\sigma_{P} \ge 3$ and $I/\sigma_{I} \ge 80$ were selected. A physical scale (black line) of $1.0$  kpc and the beam size (red circle) are shown.
\label{fig:fig8}}
\epsscale{2.}
\end{figure*}

In this manuscript, we focus on the characterization of the new observing technique and show the overall status of the observations and the polarization maps of each object. A detailed quantitative analysis of the polarization fraction and magnetic field orientation will be presented in Papers IV and V of the series. Figures \ref{fig:fig6},  \ref{fig:fig7}, and  \ref{fig:fig8} show the polarization maps of the released data. Polarization measurements are shown with a constant length to display the magnetic field orientation. The polarization measurements (E-vector) were rotated by $90^{\circ}$ to show the B-field orientation. We selected polarization measurements with $P/\sigma_P \ge 3$, $PI/\sigma_{PI} \ge 3$, $P \le 30$\%, and $I/\sigma_{I} \ge 50$ for all galaxies, except for M82, NGC~253, and NGC~2146 with  $I/\sigma_{I} \ge 80$, where $\sigma_{I}$ and $\sigma_{PI}$ are the uncertainties in Stokes $I$ and polarization intensity, respectively.

\begin{deluxetable}{lcccc}
\tablecaption{Status of released galaxy sample. \emph{Columns, from left to right:} a) Galaxy name. b) Central wavelength of the band. C) On-source requested time. D) Observed on-source time. E) completed fraction per object per band.
\label{tab:DR1} 
}
\tablecolumns{6}
\tablewidth{0pt}
\tablehead{\colhead{Galaxy} & 	\colhead{Band}  & \colhead{Requested} & \colhead{Observed} & 
\colhead{Completed$^{\star}$}  \\ 
 &		&	\colhead{Time$^{\dagger}$}	&	\colhead{Time$^{\dagger}$} 	 \\
  &	\colhead{(\um)}	&	\colhead{(h)}	&	\colhead{(h)} 	 \\
\colhead{(a)} & \colhead{(b)} & \colhead{(c)} & \colhead{(d)} & \colhead{(e)} } 
\startdata
Centaurus A &	89	&	-	    &	0.89	&	-	\\
Circinus 	&	53	&	-	    &	0.11	&	-	\\
			&	89	&	-	    &	0.89	&	-	\\
			&	214	&	-	    &	0.29	&	-	\\	
M51 		&	154	&	-	    &	2.78	&	-	\\
M82 		&	53	&	2.00	&	1.60	&	\checkmark	\\
			&	89	&	2.00	&	1.65	&	\checkmark	\\
			&	154	&	2.00	&	1.99	&	\checkmark	\\
			&	214	&	2.00	&	1.13	&	\checkmark	\\
M83 		&	154	&	6.80	&	6.60	&	\checkmark	\\
NGC~253 	&	89	&	3.00	&	1.00	&	33\% 	\\
			&	154	&	5.00	&	0.89	&	18\% 	\\
NGC~1068 	&	53	&	3.07	&	1.06	&	35\%		\\
			&	89	&	7.07	&	2.00	&	28\% 	\\
NGC~1097 	&	53	&	-	    &	1.87	&	-	\\
			&	154	&	-	    &	0.27	&	- 	\\
NGC~2146 	&	53	&	3.00	&	1.60	&	53\%		\\
			&	89	&	3.00	&	2.02	&	67\%		\\
			&	154	&	3.00	&	2.40	&	80\%		\\
			&	214	&	3.00	&	2.33	&	\checkmark	\\
NGC~3627 	&	154	&	6.80	&	4.53	&	67\%		\\
NGC~4736	&	154	&	6.80	&	2.22	&	33\%		\\
NGC~4826 	&	89	&	6.80	&	1.22	&	18\%		\\
NGC~6946 	&	154	&	6.80	&	7.20	&	\checkmark	 \\
NGC~7331 	&	154	&	6.80	&	6.10	&	\checkmark	\\	
\enddata
\tablenotetext{{\dagger}}{On-source times. Overhead of 1.08 for OTFMAP observations (Table \ref{tab:OVE}), except for M51 with a specific overhead of 2.59 for C2N observations. Observations from other SOFIA programs are labeled as `-'.}
\tablenotetext{{\star}}{the final observations are considered as completed (i.e. \checkmark) when a) completion $>85$\% or b) a desired SNR was reached. Observations from other SOFIA programs are labeled as `-'.}
\end{deluxetable}

\section{OTFMAP and C2N comparison} \label{sec:OTFvsC2N}

This section shows a quantitative comparison of the overheads and sensitivities between the C2N and OTFMAP observation modes of HAWC+.  

\subsection{Observing overheads}\label{subsec:OVE}

The OTFMAP observational strategy for polarimetry has been described in Section \ref{subsec:crush}. Here, we briefly describe the specifics of the C2N polarimetric observations using HAWC+ \citep[for further details see][]{Harper2018}. The telescope with the secondary mirror on the optical axis points to the science object. Then, the secondary mirror is moved at a frequency of $10.2$ Hz between the science object and an adjacent position on the sky. This movement is characterized by its amplitude (i.e. chop-throw) and direction (i.e. chop-angle). Each pair of science and sky images are subtracted to eliminate the background emission from the science image. To minimize the radiative offset, the telescope is moved every $30-50$s (i.e. nod time) with the same amplitude and direction as the chop. Asymmetric chop-nod is not performed with HAWC+ due to the associated large overheads. The final image is computed after a pair of subtracted chop-nod observations, which produces a positive image of the science object at the location of the boresight within the FOV of HAWC+. Chopping within the array is not performed due to the small FOVs (Table \ref{tab:HAWC}). This procedure is repeated per HWP~PA in a sequence of $5^{\circ}$, $50^{\circ}$, $27.5^{\circ}$, and $72.5^{\circ}$. As the yield of operational detectors is $\sim70$\% \citep{Harper2018}, these sets of observations are repeated in a 4-point dither sequence with a 3 pixel offset around the boresight in a given HAWC+ band. Before and after the sequence of four HWP~PAs, internal calibration images (i.e. INTCAL) of $15$s each are taken to calibrate the C2N observations and compute observing flatfields.

The total observing time is estimated using the start and end time of observations for each file, and the total on-source times are computed as follows. The total on-source time for the C2N polarimetric mode is estimated as

\begin{equation}\label{eq:C2N}
t_{\mbox{on-source,C2N}} =  \frac{\mbox{nodtime}}{2} \times n_{\mbox{dithers}} \times n_{\mbox{HWPPAs}}
\end{equation}
\noindent
where $\mbox{nodtime}$ is the nodding time where half of the time is spent on the science object, $n_{\mbox{dithers}}$ is the number of dither positions, and $n_{\mbox{HWPPAs}}$ is the number of HWP~PAs. 

The total on-source time for the OTFMAP polarimetric mode is estimated as

\begin{equation}\label{eq:OTFMAP}
t_{\mbox{on-source,OTFMAP}} = t_{\mbox{ScanDuration}} \times n_{\mbox{HWPPAs}}
\end{equation}
\noindent
where $t_{\mbox{ScanDuration}}$ is the scan duration per scan as shown in Table \ref{tab:OBS}.

We compare the observing overheads of the C2N and OTFMAP polarimetric observations of NGC~1068 and Circinus at $89$ \um. The C2N observations of NGC~1068 were presented by \citet{ELR2020} with a total observing time of $7615$s and a total on-source time of $2910$s. The chop-throw is $180$\arcsec, and the chop-angle is $90^{\circ}$ to always be along the short axis of the array. The C2N observations of Circinus were performed during the same flight, F596, as those of the OTFMAP polarimetric observations with a total observing time of $615$s and a total on-source time of $252$s. The chop-throw is $180$\arcsec, and the chop-angle is $322^{\circ}$ counter-clockwise in the East of North direction 

\begin{figure*}[ht!]
\centering
\includegraphics[angle=0,scale=0.46]{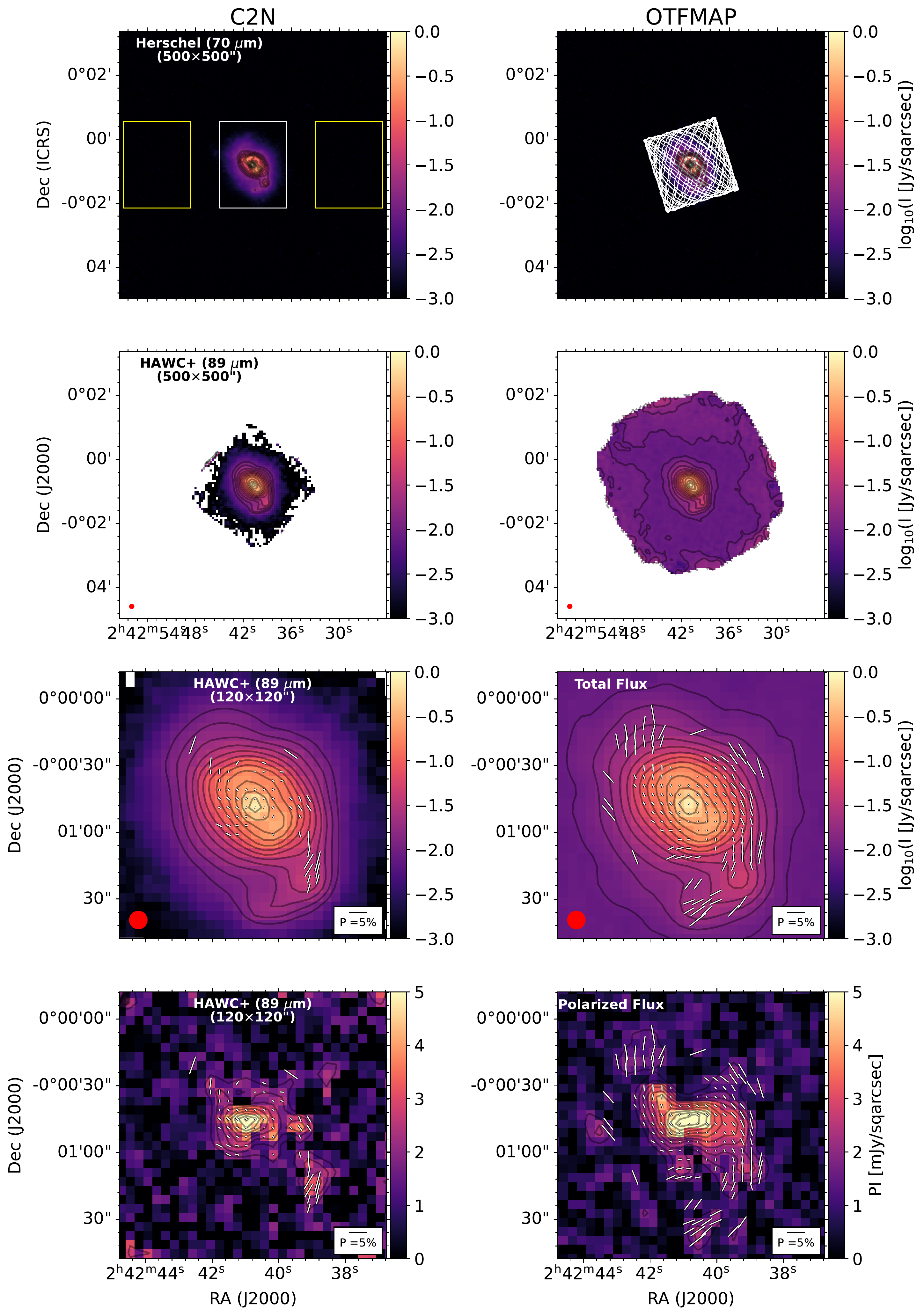}
\caption{Comparison between C2N (left) and OTFMAP (right) observations of NGC~1068 at $89$ \um. \textit{First row:} \textit{Herschel} observations at $70$ \um\ (colorscale) of NGC~1068 within a FOV of $500\times500$ sqarsec with overlaid C2N (left) and OTFMAP (right) configurations. For C2N, on-axis (white) and off-axis (yellow) positions with the FOV of HAWC+ at $89$ \um\ are shown. For OTFMAP, the Lissajous curve (white solid line) of one scan is shown. \textit{Second row:} HAWC+ total flux observations for C2N and OTFMAP within the same FOV as above. Contours start at $32\sigma_{I}$ and increase in steps of $2^{n}\sigma$, where $n=5, 5.5, 6, \dots$ and $\sigma_{I} = 0.3$ mJy/sqarcsec. The beam size is shown as a red circle in the bottom left. \textit{Third row:} Same as above within a FOV of $120\times120$ sqarcsec. Polarization measurements (white lines) are shown for $P/\sigma_{P} \ge 3$, $PI/\sigma_{PI} \ge 3$, $P<30$\%, and $I/\sigma_{I} \ge50$. A $5\%$ polarization  measurement is shown at the bottom right. \textit{Fourth row:} HAWC+ polarized flux observations (colorscale) within the same FOV and polarization as above. Contours start at $3\sigma_{PI}$ and increase in steps of $1\sigma_{PI}$, where $\sigma_{PI} = 0.46$ mJy/sqarcsec.
\label{fig:fig9}}
\epsscale{2.}
\end{figure*}

A subgroup of the OTFMAP observations presented in Table \ref{tab:OBS} were reduced to have on-source times similar to the C2N observations. OTFMAP polarimetric observations with an on-source time of $2880$s and $400$s were reduced following the steps described in Section \ref{sec:OBS} for NGC~1068 and Circinus, respectively. The associated total observing times of these subsets of observations are $3041$s and $428$s for NGC~1068 and Circinus, respectively. Figures \ref{fig:fig9} and \ref{fig:figA3} show the C2N configuration and Lissajous curves from one of the scans over the $\textit{Herschel}$ image at $70$ \um\ for NGC~1068 and Circinus, respectively.

The observing overhead factor for the C2N observing mode is estimated to be $2.53$, while an overhead of $1.08$ is estimated for the OTFMAP observing mode (Table \ref{tab:OVE}). Then, the total observing time is computed as

\begin{eqnarray}
t_{\mbox{C2N}} & = & 2.53\times t_{\mbox{on-source,C2N}} \\
t_{\mbox{OTFMAP}} &=& 1.08\times t_{\mbox{on-source,OTFMAP}}
\end{eqnarray}

The OTFMAP overhead arises from: a) the time spent to rotate the four HWP~PA, and b) the waiting time of $\sim2-4$s to ensure tracking is correctly performed between scans. The C2N overhead arises from a) chopping on the sky positions and nodding the telescope, b) INTCALs of $15$s each before and after the sequence of four dither positions, and c) the rotation of the HWP~PAs. We estimate that the OTFMAP polarimetric mode provides a total improvement in observing time overhead of a factor of $2.53/1.08 = 2.34$ with respect to the C2N polarimetric observations. The main reason for this improvement is that the galaxies are always within the FOV of the HAWC+ array during the whole integration time.

\begin{deluxetable*}{p{2.5cm}cccccc}
\tablecaption{OTFMAP vs. C2N: Overheads and sensitivities using observations of Circinus and NGC~1068 at $89$ \um. \emph{Columns, from left to right:} a) Observing mode, b) observing time overhead (i.e. total exposure time / o-source time), c) minimum intensity flux for polarimetry (MIfP) to achieve a polarization uncertainty of $0.3$\% in $900$s (including overheads), d) minimum detectable continuum polarized flux (MDCPF) for a $4\sigma$ detection in $900$s (including overheads), e) same as c) but only taken into account on-source time, f) same as d) but only taken into account on-source time.
\label{tab:OVE} 
}
\tablecolumns{6}
\tablewidth{0pt}
\tablehead{ 
\colhead{Obs. Mode} & \colhead{Overhead} & 
	\multicolumn{2}{c}{with Overheads} &	\multicolumn{2}{c}{without Overheads} \\
  &   &  
\colhead{MIfP$^{\dagger}$} 	&  \colhead{MDCPF$^{\dagger}$} &  \colhead{MIfP} 	&  \colhead{MDCPF}\\ 
     &  &
\colhead{with $\sigma_{p}=0.3$\% in 3600s} 	&  \colhead{with $4\sigma$ in 900s} &
\colhead{with $\sigma_{p}=0.3$\% in 3600s} 	&  \colhead{with $4\sigma$ in 900s} \\ 
 	&				&	\colhead{[MJy/sr]}	& \colhead{[\% Jy]} &
		\colhead{[MJy/sr]}	& \colhead{[\% Jy]}\\
\colhead{(a)} & \colhead{(b)} & \colhead{(c)} & \colhead{(d)} & \colhead{(e)} & \colhead{(f)} } 
\startdata
C2N		&	2.53	&		7010		&	59	&	4333	&	37\\
OTFMAP	&	1.08	&		2819		&	21	&	2410	&	21\\
Improvement$_{OTFMAP}^{\star}$		&	2.34	&	2.49	&	2.81	&	1.80	&	1.76 \\		
\enddata
\tablenotetext{{\star}}{Improvement of the OTFMAP shown as a ratio between OTFMAP and C2N of the quantities shown for each column.}
\tablenotetext{{\dagger}}{\citet{Harper2018} quote a MIfP and MDCPF, including overheads, of $6000$ MJy/sr and $50$ \% Jy, respectively.}
\end{deluxetable*}

\subsection{Sensitivities}\label{subsec:SEN}

The sensitivities of the total intensity mode using OTFMAP and polarimetry mode using C2N have been already characterized \citep{Harper2018}. We present the sensitivity estimations of the OTFMAP polarimetric mode at 89 \um\ (Band C) as an example of the improvement provided by this new observing mode. Further characterization in different bands will be performed and released by the SOFIA Science Center. The SOFIA observer's handbook for HAWC+\footnote{Observer's Handbook for HAWC+ can be found at \url{https://www.sofia.usra.edu/instruments/hawc}.} defines the sensitivities as follows: the Minimum Intensity flux for Polarimetry (MIfP) is given as the flux needed to reach $0.3$\% polarization uncertainty in $3600$s for an extended source; and  the Minimum Detectable Continuum Polarized Flux (MDCPF) is given as the total flux needed to reach a $4\sigma$ detection in $900$s for point sources.

\begin{figure*}[ht!]
\includegraphics[angle=0,scale=0.60]{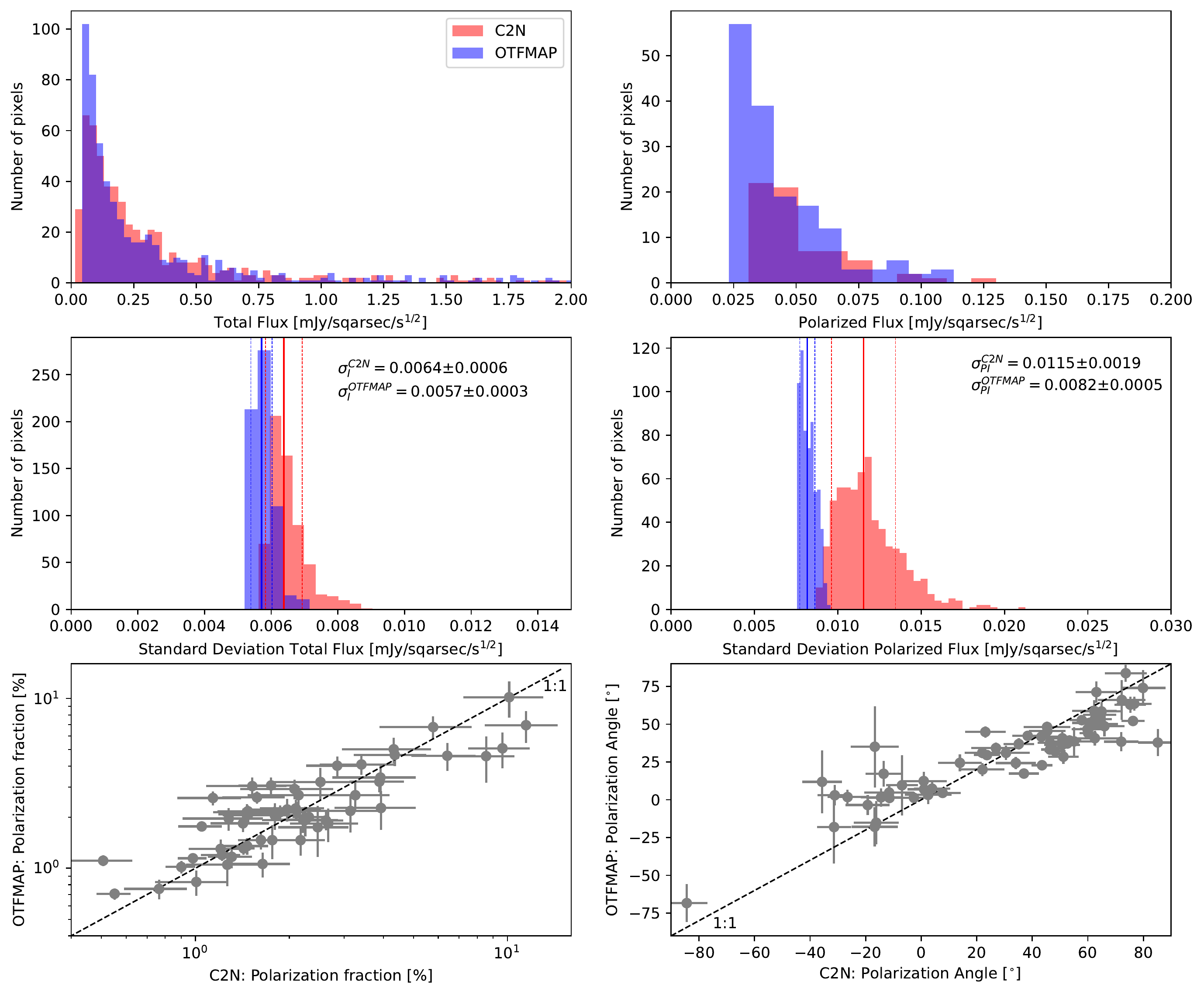}
\caption{Sensitivity comparison between C2N (red) and OTFMAP (blue) observations of NGC~1068 at $89$ \um. Histograms of the total intensity (top left), polarized intensity (top right), standard deviations of the total flux (middle left) and polarized flux (middle right) are shown. The median and $1\sigma$ uncertainty of the distribution of the standard deviations are shown in the middle panels. The polarization fraction (bottom left) and polarization angle (bottom right) for  C2N and  OTFMAP observations are shown. A 1:1 line (black dashed line) is shown.
\label{fig:fig10}}
\epsscale{2.}
\end{figure*}

Using the OTFMAP and C2N polarimetric observations of NGC~1068 and Circinus at $89$ \um, Figures \ref{fig:fig9} and \ref{fig:figA3} clearly show an increase in the number of polarization measurements with the same statistical significance (i.e. $P/\sigma_{P}\ge3$), and a larger spatial extension of the polarized flux with the same $PI/\sigma_{PI}\ge3$. Figure \ref{fig:fig10} and \ref{fig:figA4} show the histograms of the total and polarized fluxes, as well as a 1:1 plot between the polarization fraction and polarization angle for NGC~1068 and Circinus, respectively. For the 1:1 plot, measurements associated with the same WCS were plotted. We conclude that the total flux, polarized flux, polarization fraction, and position angles are in agreement between both OTFMAP and C2N polarimetric observations. In addition, after the standard deviation of the observations are normalized using their associated on-source time, we find that the standard deviation of the polarized flux decreases for the OTFMAP polarimetric observations (middle right panels of Figures \ref{fig:fig10} and \ref{fig:figA4}), demonstrating a significant improvement in sensitivity by the OTFMAP polarimetric mode.

Using the observing overheads presented in Section 3.1, we estimate a MIfP of $7010$ MJy/sr and MDCPF of $59$ \% Jy in C2N mode at $89$ \um. These values are in agreement, within the nominal $20$\% flux uncertainty, with the quoted MIfP of $6000$ MJy/sr and $50$ \% Jy by \citet{Harper2018} and the HAWC+ observer's handbook.  The estimated MIfP and MDCPF for the OTFMAP polarimetric mode are shown in Table \ref{tab:OVE}. Finally, we estimate a sensitivity improvement of $1.80$ in MIfP, without accounting for the observing overheads. After accounting for observing time overheads, the OTFMAP polarimetric mode offers total improvement in MIfP of $2.49$ times over the C2N observing mode.

\subsection{Instrumental polarization}\label{subsec:IP2}

The IP was estimated with unpolarized standard objects using C2N, OTFMAP, and polarization skydips. The IP from C2N and polarization skydips were collected in 2017 and computed by \citet{Harper2018}.  The polarization from OTFMAP of planets were presented and computed in Section \ref{subsec:IP}. Figure \ref{fig:fig5} shows the estimated IP as a percentage of the normalized Stokes $qu$ in the instrument reference frame for the C2N, OTFMAP, and polarization skydips. The numerical values and their uncertainties are shown in Table \ref{tab:IP}. We find that all three methods produce consistent and reproducible results. The uncertainties are larger in C2N and OTFMAP than in the polarization skydips due to the number of measurements used for the statistics. We take the measurements from the polarization skydips for the IP correction in our observations (Section \ref{subsec:IP}). 

We find a rotation in the IP with wavelength. The HAWC+ filters are metal grids \citep{Harper2018}, which probably introduce some level of intrinsic polarization. However, the polarization properties of these filters were not measured before they were installed in the instrument. The important result here is that the IP is reproducible and consistent throughout the different methods, and time and conditions of observations. The IP is known to arise from the tertiary mirror \citep{Harper2018}. Thus, our results show that the IP remains constant, within the uncertainties, from 2017 to 2021. This result suggests that the physical conditions of the tertiary mirror produce a negligible difference in IP during five years of observations.

\subsection{Further implementations of the OTFMAP polarization for HAWC+}\label{subsec:IMP}

Although we have shown that the OTFMAP polarimetric mode has substantially improved the overheads and sensitivities when compared to the C2N polarimetric mode, further developments on the OTFMAP polarimetric mode are still required. As mentioned above, the data processing presented in Section \ref{sec:OBS} works for objects with thermal emission within the FOV of HAWC+ in a given band. However, this procedure may not be as efficient for large-scale and diffuse thermal emission that cover areas much larger than the FOV of the HAWC+ array. This well-known deficiency of the OTFMAP observing mode can be mitigated by application of alternative OTFMAP strategies in the recovery of large-scale emission, as discussed below. 

For FIR polarimetric observations, \citet{Hildebrand2000} suggested an observing strategy involving the telescope chopping to an adjacent sky location at the same time as scanning. As in C2N, the chop is then subtracted from the closest scan in time. While this technique may remove the sky fluctuations more efficiently than the procedure we presented earlier,  it will increase the observing overheads and reduce the sensitivities as there may be a residual radiative offset after the chop subtraction. An alternative strategy consists of a continuous rotation of the HWP to modulate signal at a faster rate, usually $\sim2$ Hz, than the atmospheric fluctuations Note that HAWC+ can continuously rotate the HWP to $0.5$ Hz, providing a polarization modulation frequency of $2$ Hz \citep{Harper2018}. The time streams for each pixel can then be linearly decomposed to IP, astrophysical polarization, detector gains, detector temperature fluctuations, and atmospheric transparency fluctuations. This technique is commonly used in sub-mm polarimetric observations, for example by POL-2 on the James Clerk Maxwell Telescope \citep[JCMT;][]{WT2017},  and cosmic microwave background (CMB) experiments \citep[e.g.,][]{Johnson2007}. While this observing approach may be the most efficient, its implementation requires new  software and hardware development. These common issues are also well-known problems in radio polarimetric observations using scanning techniques. \citet{Muller2017} applied a technique called 'basket-weaving' to maps with orthogonal scanning on the sky, and \citet{Emerson1988} applied a technique called 'PLAIT' for scans oriented at any angle. Although evaluating these observing modes is outside of the scope of this manuscript, we have mentioned them as options for future implementations that could reduce the observational overheads and, as a result, SOFIA's operational cost.


\section{Conclusions}

We present the first data release (DR1) of the SOFIA Legacy program (PI: Lopez-Rodriguez, E. \& Mao, S. A.) on extragalactic magnetism. The DR1 consists of 14 galaxies, including active galactic nuclei, starbursts, and spirals, from 53 to 214 \um, which comprises 33\% (51.43h out of 155.7h) of the total awarded time of this program. We release homogeneously reduced high-level data products, ready for scientific analysis. 

We have presented the data processing of the OTFMAP polarization mode for HAWC+. The pipeline steps were applied to the newly acquired observations of galaxies from January 2020 to December 2021 within the SOFIA Legacy program on extragalactic magnetism. This new observing mode had been successfully applied to objects smaller than the FOV of the HAWC+ array in any given band. The  data processing includes: Zero-level background subtraction, PWV correction, instrumental polarization subtraction, and merging. Comparison with C2N observations show that the OTFMAP has greatly improved the polarimetric mode of HAWC+. Specifically, we estimated that OTFMAP polarimetric mode provides an improvement in observing time overhead of a factor of $2.34$, and an improvement of $1.80$ in MIfP without accounting for the observing overheads from the C2N polarimetric observations. Including these two factors, we estimate that the OTFMAP polarimetric mode offers a total improvement in MIfP of $2.49$ times over the C2N observing mode. The OFTMAP is a significant optimization of the polarimetric mode of HAWC+ as it ultimately reduces the cost of operations of SOFIA/HAWC+--more data can be collected per hour of flight, more than doubling the number of programs/papers per hour of observation.

Although the polarimetric mode of HAWC+ has been improved, we emphasize that further development of this observing mode for objects with large-scale diffuse polarized emission is still required. We briefly described several observing strategies to improve this technique for HAWC+ based on the approaches from sub-mm observations and CMB experiments.


\begin{acknowledgments}

Based on observations made with the NASA/DLR Stratospheric Observatory for Infrared Astronomy (SOFIA) under the 07\_0034, 08\_0012 Program. SOFIA is jointly operated by the Universities Space Research Association, Inc. (USRA), under NASA contract NNA17BF53C, and the Deutsches SOFIA Institut (DSI) under DLR contract 50 OK 0901 to the University of Stuttgart. KT has received funding from the European Research Council (ERC) under the European Unions Horizon 2020 research and innovation programme under grant agreement No. 771282. EN is supported by the ERC Grant ”Interstellar” (Grant agreement 740120) and has received funding from the Hellenic Foundation for Research and Innovation (H.F.R.I., project No 224). SMA is supported by the ERC Starting Grant 638707 ``Black holes and their host galaxies: co-evolution across cosmic time" and by STFC

\end{acknowledgments}

%

\vspace{5mm}
\facilities{SOFIA (HAWC+), \textit{Herschel}}


\software{
\textsc{astropy} \citep{astropy}, 
\textsc{APLpy} \citep{aplpy},
\textsc{matplotlib} \citep{hunter2007}
}

\appendix

\section{PWV correction of NGC~6946}\label{App:PWV}

\begin{figure*}[ht!]
\includegraphics[angle=0,scale=0.55]{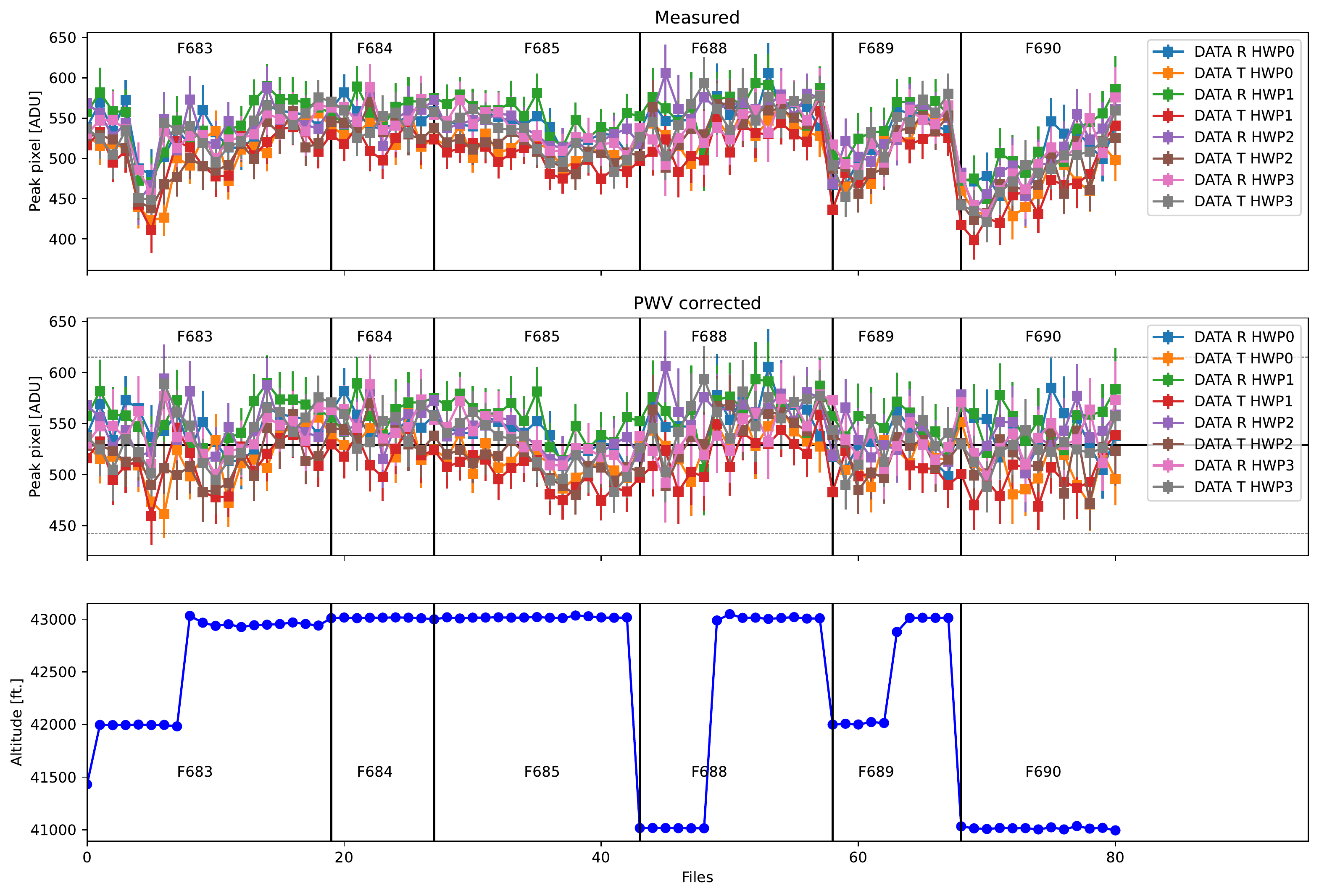}
\caption{Flux variation as a function of time and aircraft altitude for NGC6946 at 154 \um. Top panel: Measured peak fluxes of the scans per HWP~PA and $\mathcal{R}$ and $\mathcal{T}$ arrays. Middle panel: PWV corrected fluxes using the fits shown in Figure \ref{fig:figA2}. The mean (black solid line) and $2\sigma$ uncertainty (black dotted line) are shown. Bottom panel: aircraft altitude at the time of data acquisition.
\label{fig:figA1}}
\epsscale{2.}
\end{figure*}

\begin{figure}[ht!]
\center
\includegraphics[angle=0,scale=0.55]{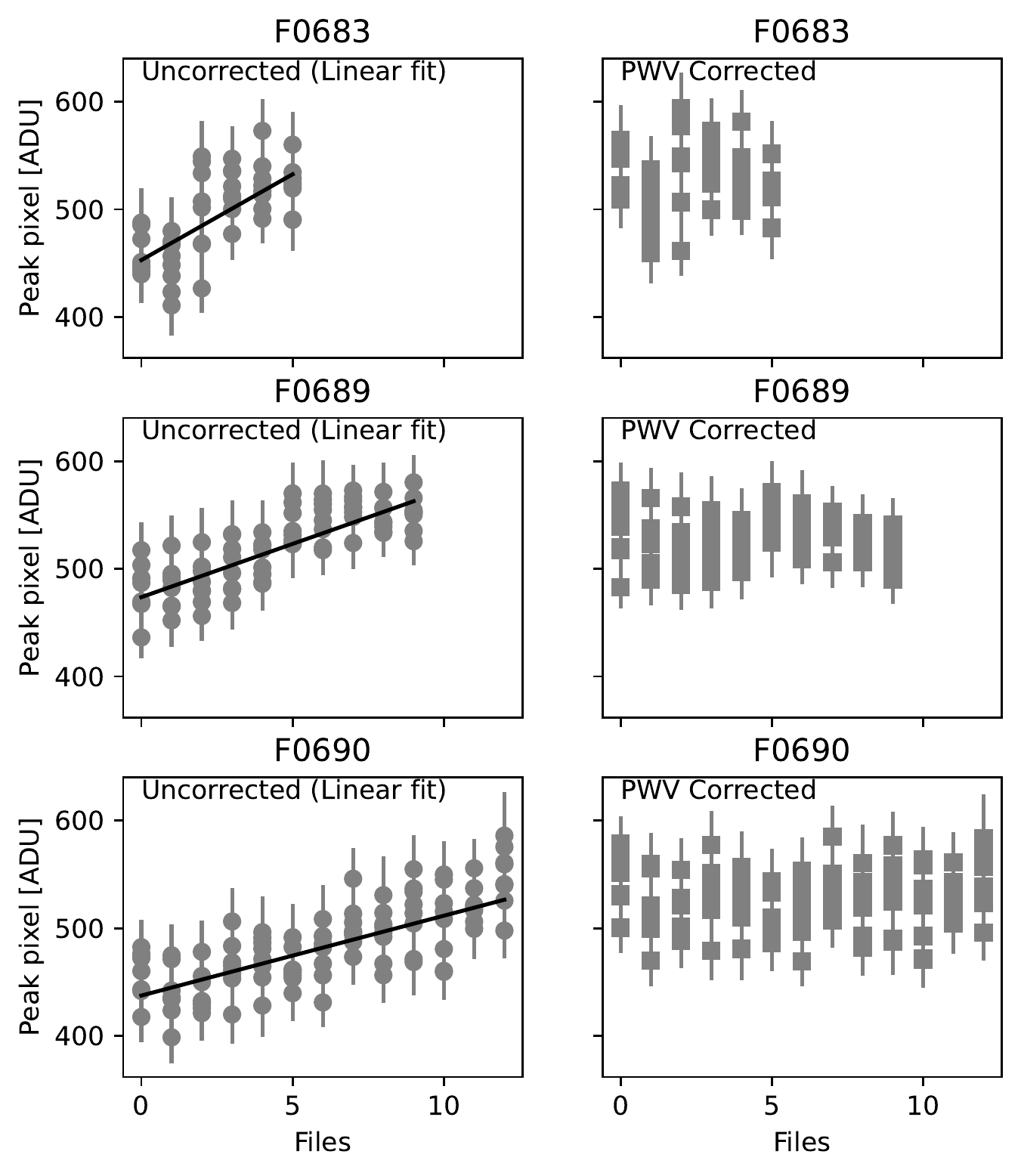}
\caption{Fluxes uncorrected and corrected by a linear fit due to PWV variations during observations of NGC~6946. Measured fluxes (left column) are fit with a linear function (black solid line) for flights F0683 (top), F0689 (middle), and F0690 (bottom). Corrected fluxes (right column) by the normalized fit to the mean flux of flights F684 and F685 are shown.
\label{fig:figA2}}
\epsscale{2.}
\end{figure}

NGC~6946 shows flux variations in several flights due to PWV fluctuations during the observations. Following the approach described  in Section \ref{subsec:PWV}, the measured fluxes (Figure \ref{fig:figA1}-top) were corrected by fitting a linear function (Figure \ref{fig:figA2}) normalized to the mean fluxes of flights F684 and F685. The corrected fluxes (Figure \ref{fig:figA1}-middle) show a flux variation of $15$\% across the full observations.

\section{Instrumental polarization}\label{App:IP}

This section shows the supporting figure described in Section \ref{subsec:IP} and \ref{subsec:IP2}.

\begin{figure}[ht!]
\centering
\includegraphics[angle=0,width=\textwidth]{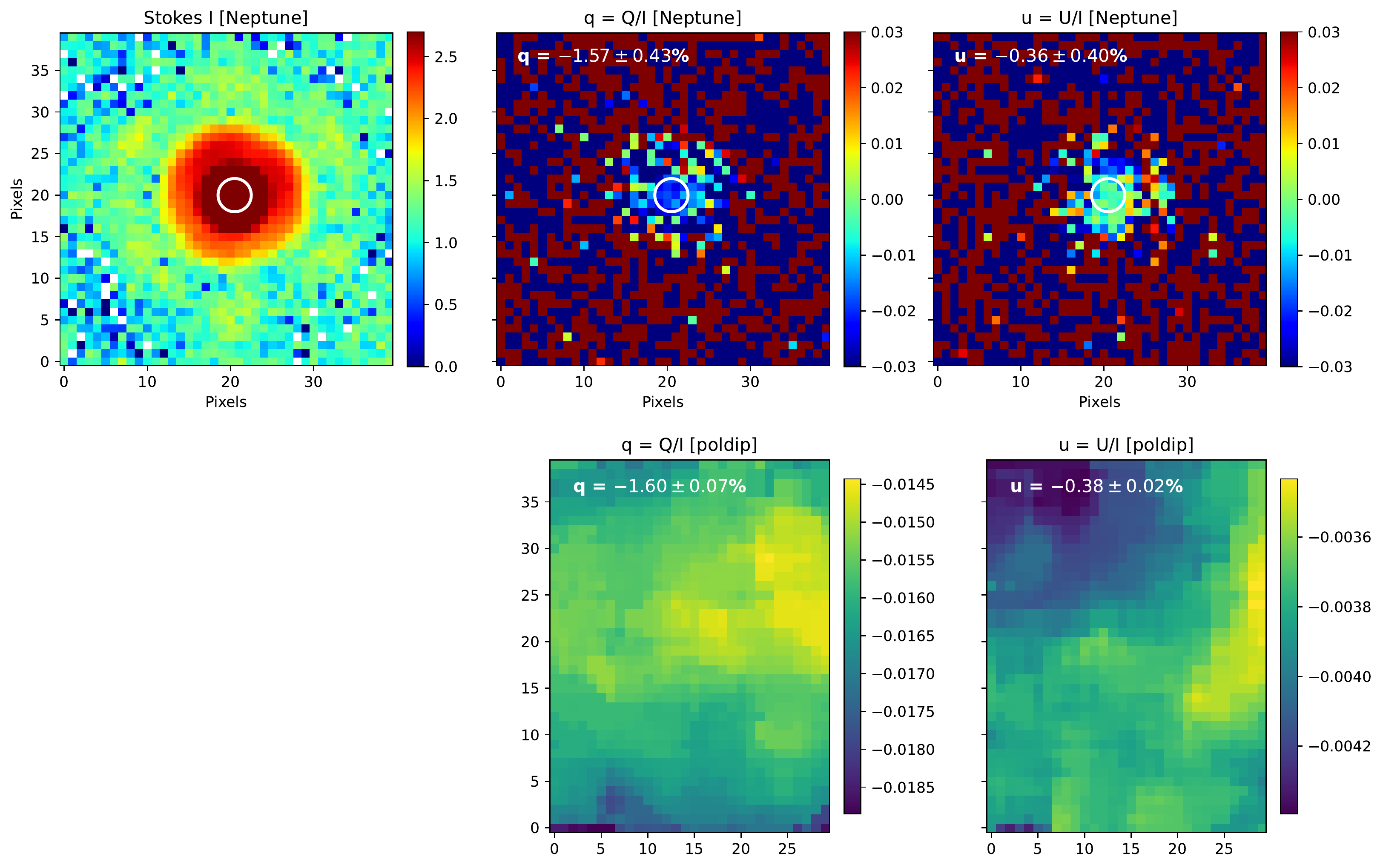}
\caption{OTFMAP polarimetric observations of a planet (Neptune) and polarization skydip (poldip) at $53$ \um. 
Top panels: Polarization observations of a planet. Stokes I (left) in $\log_{10}$ scale, normalized Stokes $q=Q/U$ (middle) and $u=U/I$ (right) are shown. The FOV is $40\times40$ pixels ($48\times48$ sqarcsec, with pixelscale of $1\farcs21$). 
Bottom panels: Polarization observations of a skydip. Normalized Stokes $q=Q/U$ (middle) and $u/I$ (right) are shown. The FOV is $32\times40$ pixels ($82\times102$ sqarcsec, with a pixelsxale of $2\farcs55$).
Definitions and details of the estimation of the IP are described in Section \ref{subsec:IP} and \ref{subsec:IP2}.
\label{fig:figA3n}}
\epsscale{2.}
\end{figure}

\section{OTFMAP vs. C2N for Circinus}\label{app:OTFvsC2N}

\begin{figure}[ht!]
\center
\includegraphics[angle=0,scale=0.45]{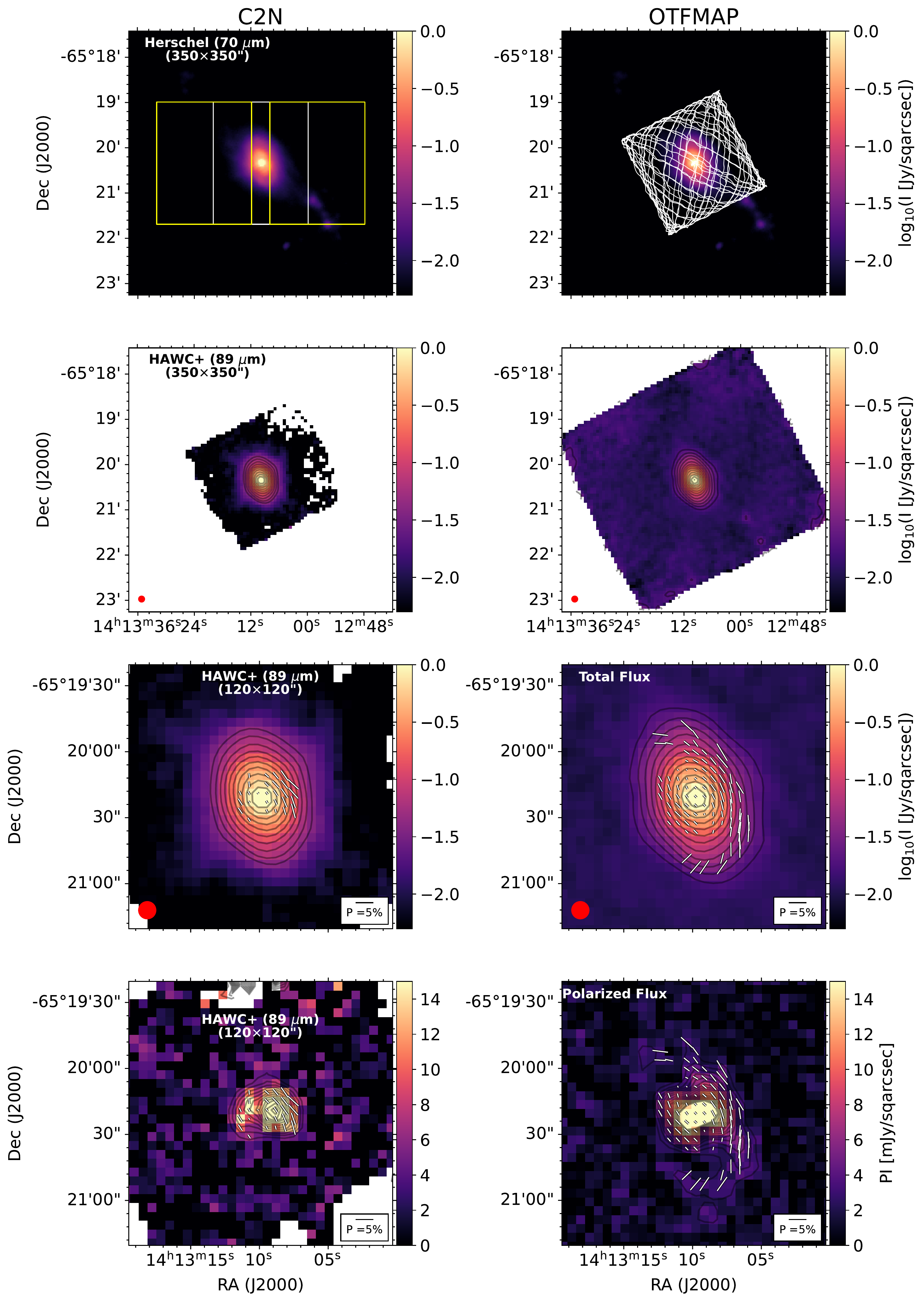}
\caption{Comparison between C2N (left) and OTFMAP (right) observations of Circinus at $89$ \um. \textit{First row:} \textit{Herschel} observations at $70$ \um\ (colorscale) of Circinus within a FOV of $350\times350$ sqarsec with overlaid C2N (left) and OTFMAP (right) configurations. For C2N, on-source (white) and off-source (yellow) positions with the FOV of HAWC+ at $89$ \um\ are shown. For OTFMAP, the Lissajous curve (white solid line) of the scan is shown. \textit{Second row:} HAWC+ total flux observations for C2N and OTFMAP within the same FOV as above. Contours start at $32\sigma_{I}$ and increase in steps of $2^{n}\sigma$, where $n=5, 5.5, 6, \dots$ and $\sigma_{I} = 0.3$ mJy/sqarcsec. The beam size is shown as a red circle in the bottom left. \textit{Third row:} Same as above within a FOV of $120\times120$ sqarcsec. Polarization measurements (white lines) are shown for $P/\sigma_{P} \ge 3$, $PI/\sigma_{PI} \ge 3$, $P<30$\%, and $I/\sigma_{I} \ge50$. A $5\%$ polarization  measurement is shown at the bottom right. \textit{Fourth row:} HAWC+ polarized flux observations (colorscale) within the same FOV and polarization as above. Contours start at $3\sigma_{PI}$ and increase in steps of $1\sigma_{PI}$, where $\sigma_{PI} = 0.45$ mJy/sqarcsec.
\label{fig:figA3}}
\epsscale{2.}
\end{figure}

\begin{figure}[ht!]
\center
\includegraphics[angle=0,scale=0.55]{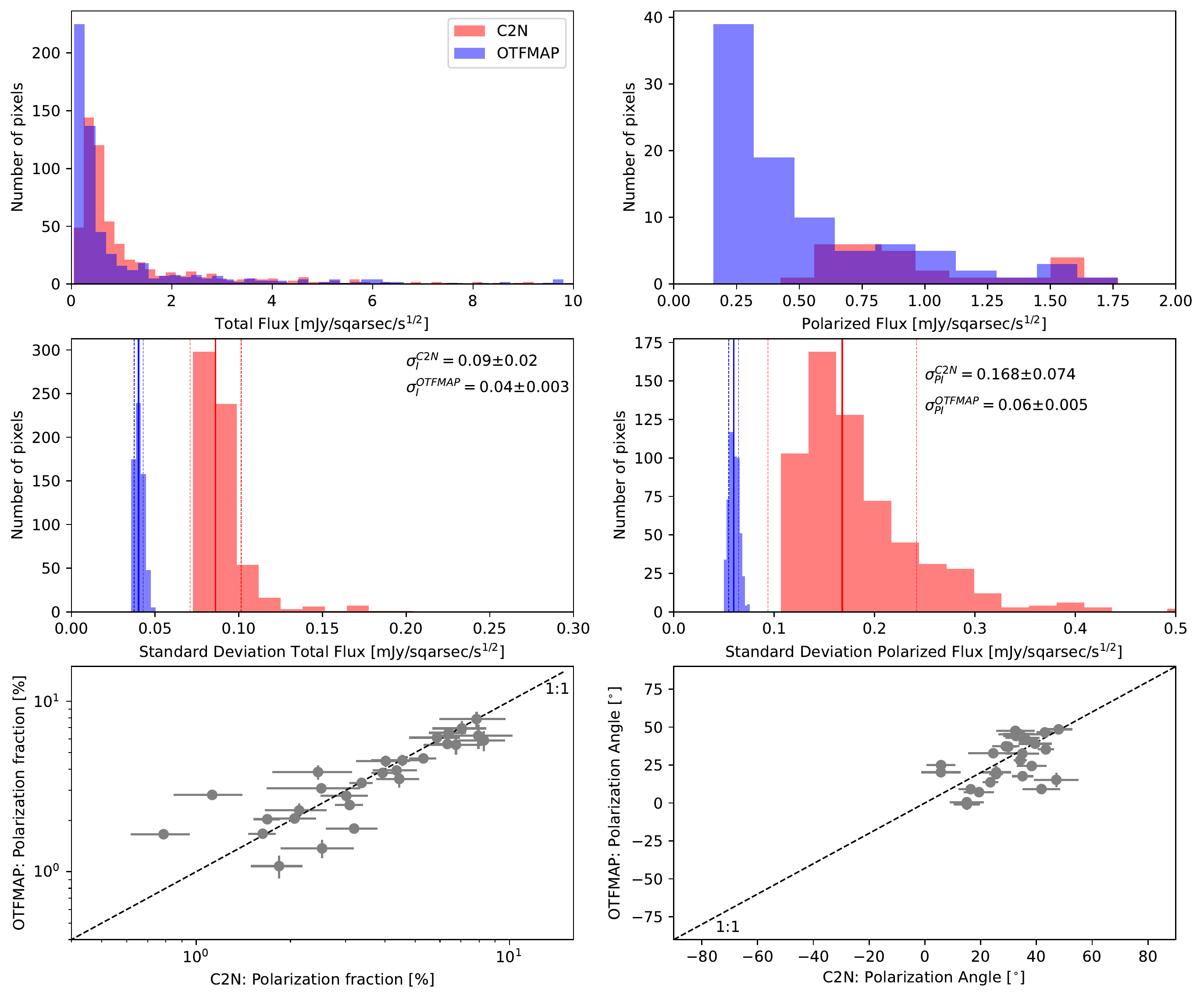}
\caption{Sensitivity comparison between C2N (red) and OTFMAP (blue) observations of Circinus at $89$ \um. Histograms of the total intensity (top left), polarized intensity (top right), standard deviations of the total flux (middle left) and polarized flux (middle right) are shown. The median and $1\sigma$ uncertainty of the distribution of the standard deviations are shown in the middle panels. The polarization fraction (bottom left) and polarization angle (bottom right) for  C2N and  OTFMAP observations are shown. A 1:1 line (black dashed line) is shown.
\label{fig:figA4}}
\epsscale{2.}
\end{figure}

OTFMAP and C2N comparisons for the observations of Circinus at 89 \um. Figure \ref{fig:figA3} shows the instrumental configuration and polarization maps for both observing modes. Figure \ref{fig:figA4} presents the histograms of the total and polarized flux, and their standard deviations. The 1:1 plots of the polarization fraction and polarization angle are shown.

\bibliography{references}

\begin{thebibliography}{}
\expandafter\ifx\csname natexlab\endcsname\relax\def\natexlab#1{#1}\fi
\providecommand{\url}[1]{\href{#1}{#1}}
\providecommand{\dodoi}[1]{doi:~\href{http://doi.org/#1}{\nolinkurl{#1}}}
\providecommand{\doeprint}[1]{\href{http://ascl.net/#1}{\nolinkurl{http://ascl.net/#1}}}
\providecommand{\doarXiv}[1]{\href{https://arxiv.org/abs/#1}{\nolinkurl{https://arxiv.org/abs/#1}}}

\bibitem[{{Astropy Collaboration} {et~al.}(2013){Astropy Collaboration},
  {Robitaille}, {Tollerud}, {Greenfield}, {Droettboom}, {Bray}, {Aldcroft},
  {Davis}, {Ginsburg}, {Price-Whelan}, {Kerzendorf}, {Conley}, {Crighton},
  {Barbary}, {Muna}, {Ferguson}, {Grollier}, {Parikh}, {Nair}, {Unther},
  {Deil}, {Woillez}, {Conseil}, {Kramer}, {Turner}, {Singer}, {Fox}, {Weaver},
  {Zabalza}, {Edwards}, {Azalee Bostroem}, {Burke}, {Casey}, {Crawford},
  {Dencheva}, {Ely}, {Jenness}, {Labrie}, {Lim}, {Pierfederici}, {Pontzen},
  {Ptak}, {Refsdal}, {Servillat}, \& {Streicher}}]{astropy}
{Astropy Collaboration}, {Robitaille}, T.~P., {Tollerud}, E.~J., {et~al.} 2013,
  \aap, 558, A33, \dodoi{10.1051/0004-6361/201322068}

\bibitem[{{Borlaff} {et~al.}(2021){Borlaff}, {Lopez-Rodriguez}, {Beck},
  {Stepanov}, {Ntormousi}, {Hughes}, {Tassis}, {Marcum}, {Grosset}, {Beckman},
  {Proudfit}, {Clark}, {D{\'\i}az-Santos}, {Mao}, {Reach}, {Roman-Duval},
  {Subramanian}, {Tram}, {Zweibel}, \& {SOFIA Legacy Team}}]{Borlaff2021}
{Borlaff}, A.~S., {Lopez-Rodriguez}, E., {Beck}, R., {et~al.} 2021, arXiv
  e-prints, arXiv:2105.09315.
\newblock \doarXiv{2105.09315}

\bibitem[{{Burtscher} {et~al.}(2020){Burtscher}, {Politopoulos},
  {Fern{\'a}ndez-Acosta}, {Agocs}, {van den Ancker}, {van Boekel}, {Brandl},
  {K{\"a}ufl}, {Pantin}, {Pietrow}, {Siebenmorgen}, {Stuik}, {Tristram}, \& {de
  Wit}}]{Burtscher2020}
{Burtscher}, L., {Politopoulos}, I., {Fern{\'a}ndez-Acosta}, S., {et~al.} 2020,
  in Society of Photo-Optical Instrumentation Engineers (SPIE) Conference
  Series, Vol. 11447, Society of Photo-Optical Instrumentation Engineers (SPIE)
  Conference Series, 114477L, \dodoi{10.1117/12.2576271}

\bibitem[{{Cantalupo} {et~al.}(2010){Cantalupo}, {Borrill}, {Jaffe}, {Kisner},
  \& {Stompor}}]{Cantalupo2010}
{Cantalupo}, C.~M., {Borrill}, J.~D., {Jaffe}, A.~H., {Kisner}, T.~S., \&
  {Stompor}, R. 2010, \apjs, 187, 212, \dodoi{10.1088/0067-0049/187/1/212}

\bibitem[{{Chapin} {et~al.}(2013){Chapin}, {Berry}, {Gibb}, {Jenness}, {Scott},
  {Tilanus}, {Economou}, \& {Holland}}]{Chapin2013}
{Chapin}, E.~L., {Berry}, D.~S., {Gibb}, A.~G., {et~al.} 2013, \mnras, 430,
  2545, \dodoi{10.1093/mnras/stt052}

\bibitem[{{Dowell} {et~al.}(2010){Dowell}, {Cook}, {Harper}, {Lin}, {Looney},
  {Novak}, {Stephens}, {Berthoud}, {Chuss}, {Crutcher}, {Dotson}, {Hildebrand
  }, {Houde}, {Jones}, {Krejny}, {Lazarian}, {Moseley}, {Tassis},
  {Vaillancourt}, \& {Werner}}]{Dowell2010}
{Dowell}, C.~D., {Cook}, B.~T., {Harper}, D.~A., {et~al.} 2010, in Society of
  Photo-Optical Instrumentation Engineers (SPIE) Conference Series, Vol. 7735,
  \procspie, 77356H, \dodoi{10.1117/12.857842}

\bibitem[{{Emerson} \& {Graeve}(1988)}]{Emerson1988}
{Emerson}, D.~T., \& {Graeve}, R. 1988, \aap, 190, 353

\bibitem[{{Gordon} {et~al.}(2018){Gordon}, {Lopez-Rodriguez}, {Andersson},
  {Clarke}, {Coude}, {Moullet}, {Richards}, {Shuping}, {Vacca}, \&
  {Yorke}}]{Gordon2018}
{Gordon}, M.~S., {Lopez-Rodriguez}, E., {Andersson}, B.~G., {et~al.} 2018,
  arXiv e-prints, arXiv:1811.03100.
\newblock \doarXiv{1811.03100}

\bibitem[{{Harper} {et~al.}(2018){Harper}, {Runyan}, {Dowell}, {Wirth},
  {Amato}, {Ames}, {Amiri}, {Banks}, {Bartels}, {Benford}, {Berthoud},
  {Buchanan}, {Casey}, {Chapman}, {Chuss}, {Cook}, {Derro}, {Dotson}, {Evans},
  {Fixsen}, {Gatley}, {Guerra}, {Halpern}, {Hamilton}, {Hamlin}, {Hansen},
  {Heimsath}, {Hermida}, {Hilton}, {Hirsch}, {Hollister}, {Hostetter}, {Irwin},
  {Jhabvala}, {Jhabvala}, {Kastner}, {Kov{\'a}cs}, {Lin}, {Loewenstein},
  {Looney}, {Lopez-Rodriguez}, {Maher}, {Michail}, {Miller}, {Moseley},
  {Novak}, {Pernic}, {Rennick}, {Rhody}, {Sandberg}, {Sand ford}, {Santos},
  {Shafer}, {Sharp}, {Shirron}, {Siah}, {Silverberg}, {Sparr}, {Spotz},
  {Staguhn}, {Toorian}, {Towey}, {Tuttle}, {Vaillancourt}, {Voellmer},
  {Volpert}, {Wang}, \& {Wollack}}]{Harper2018}
{Harper}, D.~A., {Runyan}, M.~C., {Dowell}, C.~D., {et~al.} 2018, Journal of
  Astronomical Instrumentation, 7, 1840008, \dodoi{10.1142/S2251171718400081}

\bibitem[{{Haslam}(1974)}]{Haslam1974}
{Haslam}, C.~G.~T. 1974, \aaps, 15, 333

\bibitem[{{Hildebrand} {et~al.}(2000){Hildebrand}, {Davidson}, {Dotson},
  {Dowell}, {Novak}, \& {Vaillancourt}}]{Hildebrand2000}
{Hildebrand}, R.~H., {Davidson}, J.~A., {Dotson}, J.~L., {et~al.} 2000, \pasp,
  112, 1215, \dodoi{10.1086/316613}

\bibitem[{{Hunter}(2007)}]{hunter2007}
{Hunter}, J.~D. 2007, Computing in Science \& Engineering, 9, 3

\bibitem[{{Johnson} {et~al.}(2007){Johnson}, {Collins}, {Abroe}, {Ade}, {Bock},
  {Borrill}, {Boscaleri}, {de Bernardis}, {Hanany}, {Jaffe}, {Jones}, {Lee},
  {Levinson}, {Matsumura}, {Rabii}, {Renbarger}, {Richards}, {Smoot},
  {Stompor}, {Tran}, {Winant}, {Wu}, \& {Zuntz}}]{Johnson2007}
{Johnson}, B.~R., {Collins}, J., {Abroe}, M.~E., {et~al.} 2007, \apj, 665, 42,
  \dodoi{10.1086/518105}

\bibitem[{{Kov{\'a}cs}(2006)}]{kovacs2006}
{Kov{\'a}cs}, A. 2006, PhD thesis, Caltech

\bibitem[{{Kov{\'a}cs}(2008{\natexlab{a}})}]{Kovacs2008b}
{Kov{\'a}cs}, A. 2008{\natexlab{a}}, in Society of Photo-Optical
  Instrumentation Engineers (SPIE) Conference Series, Vol. 7020, Millimeter and
  Submillimeter Detectors and Instrumentation for Astronomy IV, ed. W.~D.
  {Duncan}, W.~S. {Holland}, S.~{Withington}, \& J.~{Zmuidzinas}, 702007,
  \dodoi{10.1117/12.790272}

\bibitem[{{Kov{\'a}cs}(2008{\natexlab{b}})}]{kovacs2008}
{Kov{\'a}cs}, A. 2008{\natexlab{b}}, in Society of Photo-Optical
  Instrumentation Engineers (SPIE) Conference Series, Vol. 7020, \procspie,
  70201S, \dodoi{10.1117/12.790276}

\bibitem[{{Li} {et~al.}(2021){Li}, {Lopez-Rodriguez}, {Ajeddig}, {Andr{\'e}},
  {McKee}, {Rho}, \& {Klein}}]{Li2021}
{Li}, P.~S., {Lopez-Rodriguez}, E., {Ajeddig}, H., {et~al.} 2021, \mnras,
  \dodoi{10.1093/mnras/stab3448}

\bibitem[{{Lopez-Rodriguez}(2021)}]{ELR2021}
{Lopez-Rodriguez}, E. 2021, Nature Astronomy,
  \dodoi{10.1038/s41550-021-01329-9}

\bibitem[{{Lopez-Rodriguez} {et~al.}(2020){Lopez-Rodriguez}, {Dowell}, {Jones},
  {Harper}, {Berthoud}, {Chuss}, {Dale}, {Guerra}, {Hamilton}, {Looney},
  {Michail}, {Nikutta}, {Novak}, {Santos}, {Sheth}, {Siah}, {Staguhn},
  {Stephens}, {Tassis}, {Trinh}, {Ward-Thompson}, {Werner}, {Wollack},
  {Zweibel}, \& {HAWC+Science Team}}]{ELR2020}
{Lopez-Rodriguez}, E., {Dowell}, C.~D., {Jones}, T.~J., {et~al.} 2020, \apj,
  888, 66, \dodoi{10.3847/1538-4357/ab5849}

\bibitem[{{Lopez-Rodriguez} {et~al.}(2021){Lopez-Rodriguez}, {Beck}, {Clark},
  {Hughes}, {Borlaff}, {Ntormousi}, {Grosset}, {Tassis}, {Beckman},
  {Subramanian}, {Dale}, \& {D{\'\i}az-Santos}}]{ELR2021c}
{Lopez-Rodriguez}, E., {Beck}, R., {Clark}, S.~E., {et~al.} 2021, \apj, 923,
  150, \dodoi{10.3847/1538-4357/ac2e01}

\bibitem[{{M{\"u}ller} {et~al.}(2017){M{\"u}ller}, {Krause}, {Beck}, \&
  {Schmidt}}]{Muller2017}
{M{\"u}ller}, P., {Krause}, M., {Beck}, R., \& {Schmidt}, P. 2017, \aap, 606,
  A41, \dodoi{10.1051/0004-6361/201731257}

\bibitem[{{Ohsawa} {et~al.}(2018){Ohsawa}, {Sako}, {Miyata}, {Kamizuka},
  {Okada}, {Mori}, {Uchiyama}, {Yamaguchi}, {Fujiyoshi}, {Morii}, \&
  {Ikeda}}]{Ohsawa2018}
{Ohsawa}, R., {Sako}, S., {Miyata}, T., {et~al.} 2018, \apj, 857, 37,
  \dodoi{10.3847/1538-4357/aab6ae}

\bibitem[{{Patanchon} {et~al.}(2008){Patanchon}, {Ade}, {Bock}, {Chapin},
  {Devlin}, {Dicker}, {Griffin}, {Gundersen}, {Halpern}, {Hargrave}, {Hughes},
  {Klein}, {Marsden}, {Martin}, {Mauskopf}, {Netterfield}, {Olmi}, {Pascale},
  {Rex}, {Scott}, {Semisch}, {Truch}, {Tucker}, {Tucker}, {Viero}, \&
  {Wiebe}}]{Patanchon2008}
{Patanchon}, G., {Ade}, P.~A.~R., {Bock}, J.~J., {et~al.} 2008, \apj, 681, 708,
  \dodoi{10.1086/588543}

\bibitem[{{Reichertz} {et~al.}(2001){Reichertz}, {Weferling}, {Esch}, \&
  {Kreysa}}]{Reichertz2001}
{Reichertz}, L.~A., {Weferling}, B., {Esch}, W., \& {Kreysa}, E. 2001, \aap,
  379, 735, \dodoi{10.1051/0004-6361:20011227}

\bibitem[{{Robitaille} \& {Bressert}(2012)}]{aplpy}
{Robitaille}, T., \& {Bressert}, E. 2012, {APLpy: Astronomical Plotting Library
  in Python}.
\newblock \doeprint{1208.017}

\bibitem[{{Roussel}(2013)}]{Roussel2013}
{Roussel}, H. 2013, \pasp, 125, 1126, \dodoi{10.1086/673310}

\bibitem[{{Tegmark}(1997)}]{Tegmark1997}
{Tegmark}, M. 1997, \prd, 56, 4514, \dodoi{10.1103/PhysRevD.56.4514}

\bibitem[{{Vaillancourt} {et~al.}(2007){Vaillancourt}, {Chuss}, {Crutcher},
  {Dotson}, {Dowell}, {Harper}, {Hildebrand}, {Jones}, {Lazarian}, {Novak}, \&
  {Werner}}]{Vaillancourt2007}
{Vaillancourt}, J.~E., {Chuss}, D.~T., {Crutcher}, R.~M., {et~al.} 2007, in
  Society of Photo-Optical Instrumentation Engineers (SPIE) Conference Series,
  Vol. 6678, \procspie, 66780D, \dodoi{10.1117/12.730922}

\bibitem[{{Ward-Thompson} {et~al.}(2017){Ward-Thompson}, {Pattle}, {Bastien},
  {Furuya}, {Kwon}, {Lai}, {Qiu}, {Berry}, {Choi}, {Coud{\'e}}, {Di Francesco},
  {Hoang}, {Franzmann}, {Friberg}, {Graves}, {Greaves}, {Houde}, {Johnstone},
  {Kirk}, {Koch}, {Kwon}, {Lee}, {Li}, {Matthews}, {Mottram}, {Parsons}, {Pon},
  {Rao}, {Rawlings}, {Shinnaga}, {Sadavoy}, {van Loo}, {Aso}, {Byun},
  {Eswaraiah}, {Chen}, {Chen}, {Chen}, {Ching}, {Cho}, {Chrysostomou}, {Chung},
  {Doi}, {Drabek-Maunder}, {Eyres}, {Fiege}, {Friesen}, {Fuller}, {Gledhill},
  {Griffin}, {Gu}, {Hasegawa}, {Hatchell}, {Hayashi}, {Holland}, {Inoue},
  {Inutsuka}, {Iwasaki}, {Jeong}, {Kang}, {Kang}, {Kang}, {Kawabata}, {Kemper},
  {Kim}, {Kim}, {Kim}, {Kim}, {Kim}, {Kim}, {Lacaille}, {Lee}, {Lee}, {Li},
  {Li}, {Liu}, {Liu}, {Liu}, {Liu}, {Lyo}, {Mairs}, {Matsumura},
  {Moriarty-Schieven}, {Nakamura}, {Nakanishi}, {Ohashi}, {Onaka}, {Peretto},
  {Pyo}, {Qian}, {Retter}, {Richer}, {Rigby}, {Robitaille}, {Savini}, {Scaife},
  {Soam}, {Tamura}, {Tang}, {Tomisaka}, {Wang}, {Wang}, {Whitworth}, {Yen},
  {Yoo}, {Yuan}, {Zhang}, {Zhang}, {Zhou}, {Zhu}, {Andr{\'e}}, {Dowell},
  {Falle}, \& {Tsukamoto}}]{WT2017}
{Ward-Thompson}, D., {Pattle}, K., {Bastien}, P., {et~al.} 2017, \apj, 842, 66,
  \dodoi{10.3847/1538-4357/aa70a0}

\bibitem[{{Waskett} {et~al.}(2007){Waskett}, {Sibthorpe}, {Griffin}, \&
  {Chanial}}]{Waskett2007}
{Waskett}, T.~J., {Sibthorpe}, B., {Griffin}, M.~J., \& {Chanial}, P.~F. 2007,
  \mnras, 381, 1583, \dodoi{10.1111/j.1365-2966.2007.12327.x}

\bibitem[{{Weferling} {et~al.}(2002){Weferling}, {Reichertz}, {Schmid-Burgk},
  \& {Kreysa}}]{Weferling2002}
{Weferling}, B., {Reichertz}, L.~A., {Schmid-Burgk}, J., \& {Kreysa}, E. 2002,
  \aap, 383, 1088, \dodoi{10.1051/0004-6361:20011617}

\end{thebibliography}



\end{document}